\documentclass[journal]{IEEEtran}
\usepackage{amsmath,graphicx}
\usepackage{amssymb}
\usepackage{lineno,hyperref}
\usepackage{url}
\usepackage{float}
\usepackage{textcomp}
\usepackage{xcolor}
\usepackage{algorithm}
\usepackage{subcaption}
\usepackage{algpseudocode}
\usepackage{lipsum}
\usepackage{multirow}
\usepackage{enumitem}

\begin{document}

\title{Adaptive Joint Estimation of \\ Temporal Vertex and Edge Signals}
% \author{Yi~Yan,~Tian~Xie,~and~Ercan~E.~Kuruoglu
\author{Yi~Yan,~\IEEEmembership{Student~Member,~IEEE,}~Tian~Xie,~and~Ercan~E.~Kuruoglu,~\IEEEmembership{Senior~Member,~IEEE}
\thanks{Yi Yan and Ercan E. Kuruoglu are affiliated with Tsinghua-Berkeley Shenzhen Institute, Shenzhen International Graduate School, Tsinghua University, Shenzhen, China; Tian Xie was with the Department of Electrical and Computer Engineering, University of Southern California. This work was supported by the Tsinghua Shenzhen International Graduate School Startup Fund under Grant QD2022024C and the Shenzhen Science and Technology Innovation Commission under Grant JCYJ20220530143002005. (Corresponding author: E. E. Kuruoglu. emails: y-yan20@mails.tsinghua.edu.cn; kuruoglu@sz.tsinghua.edu.cn; tianxie143@gmail.com)}
% \thanks{Manuscript received March XX, 2024; revised XXX. XX, 2024.}
}

% \markboth{IEEE TRANSACTIONS ON SIGNAL AND INFORMATION PROCESSING OVER NETWORKS,~Vol.~XX, No.~XX, XXX.~2024}%
% {Yan \MakeLowercase{\textit{et al.}}: Adaptive Joint Estimation of Temporal Vertex and Edge Signals}

\maketitle

\begin{abstract}
The adaptive estimation of coexisting temporal vertex (node) and edge signals on graphs is a critical task when a change in edge signals influences the temporal dynamics of the vertex signals. However, the current Graph Signal Processing algorithms mostly consider only the signals existing on the graph vertices and have neglected the fact that signals can reside on the edges. We propose an Adaptive Joint Vertex-Edge Estimation (AJVEE) algorithm for jointly estimating time-varying vertex and edge signals through a time-varying regression, incorporating both vertex signal filtering and edge signal filtering. Accompanying AJVEE is a newly proposed Adaptive Least Mean Square procedure based on the Hodge Laplacian (ALMS-Hodge), which is inspired by classical adaptive filters combining simplicial filtering and simplicial regression. AJVEE is able to operate jointly on the vertices and edges by merging two ALMS-Hodge specified on the vertices and edges into a unified formulation. A more generalized case extending AJVEE beyond the vertices and edges is being discussed. Experimenting on real-world traffic networks and population mobility networks, we have confirmed that our proposed AJVEE algorithm could accurately and jointly track time-varying vertex and edge signals on graphs.
\end{abstract}


\IEEEpeerreviewmaketitle

\section{Introduction}\label{sec_intro}
\IEEEPARstart{G}{raphs} have recently become a popular and impactful research topic due to their effective ability to represent interacting multivariate signals.
However, several challenges must be addressed to enhance the effectiveness of signal processing algorithms on multivariate time-varying signals that reside on graph-like topological structures. 
These challenges include representing topological irregularity, estimating unknown data from known data, removing noise present in the data, and extracting cross-space-time variations.
In graph signal processing (GSP), graphs have demonstrated their capacity to represent information in various real-world scenarios and disciplines of studies. 
Graphs have been studied extensively and attracted considerable attention due to their irregular structure, as well as their effectiveness in representing interactions among data \cite{Sandryhaila_2014_big_data, Shuman_2013_emerging, Ortega_2018}. 
Measurements taken by multiple sensors could be used to construct sensor graphs based on the locations of the sensors, with the data features embedded on the vertices.
Examples of sensor graphs include recordings of 5G signal reception strength graph \cite{bib_LMS}, air-quality graph \cite{bib_GCN_air}, and nationwide temperature graph \cite{bib_NLMS}. 
In bioinformatics, biological networks can model time-varying gene interactions. \cite{Kuruoglu_2016}, functional brain imaging \cite{Huang_2018}, and biomedicine \cite{li_2022_graph}. 
Graphical models are proposed in finance to reflect the interplay of the stock market from the interaction among stocks \cite{Yang_2020_estimating}. 

Despite the success of GSP, GSP solely considers data on the graph vertices, meaning that higher-order signals, such as signals on graph edges, are not being represented. 
To provide a few examples, traffic data on road networks recorded on the edges \cite{Jia_2019, han_2019_computing}, gene interaction over time forms a set of time-varying graph edges \cite{Kuruoglu_2016}, multi-agent systems use time-varying edge data to model the agent interactions \cite{yang_2021_Adaptive}. 
A citation complex can be viewed as a higher-ordered social network and formed from constructing simplices among co-author collaborations: the number of citations of a 2-author is a signal on the graph edges and the number of citations of a 3-author paper is a signal on a triangle \cite{Yang_2022}. 
The spread of COVID-19 across multiple regions can be represented using a series of dynamic graphs, where the vertices record the confirmed cases in each region, and the edges represent population mobility between regions.
In existing GSP algorithms that process time-varying vertex signals, the interaction among vertices is modeled using fixed parameters at all time instances \cite{Hong_GGARCH_2023}. 
Failing to capture the time-varying interaction would lead to an inaccurate estimation. 

The recently emerged Topological Signal Processing (TSP) has been established as an alternative point of view from GSP for higher-ordered signals, overcoming the limitation of GSP by generalizing graphs as simplicial complexes. 
The signals on the vertices of a graph are considered as signals of the $0$ order simplex, while higher order simplices could be used to represent signals on the edges and beyond \cite{Barbarossa_2020, Barbarossa_2020_SPM}. 
In \cite{Jia_2019}, semi-supervised learning for data on the graph edges is done using the Hodge Laplacians and flow conservation. 
An analogy of spectral wavelets on simplicial complexes called Hodgelets is proposed in \cite{Roddenberry_2022}. 
The work of \cite{Battiloro_2022_Topological} further proposed topological Slepians, which are the signals concentrated on the topological domain and localized on the frequency domain. 
In \cite{Yang_2021_FIR, Sardellitti_2022}, finite impulse response filters on simplicial complexes were discussed, and the work of \cite{Yang_2022_Simplicial} further defined the convolution operation for signals on a simplicial complex. 
Recently, the proposal of using a Dirac operator allows the merging of fixed time-invariant signals from different orders of simplex \cite{calmon2023_higher_order_dirac}. 
Chebyshev polynomial approximations of spectral filters on simplicial complexes are provided in \cite{Yang_2022_Simplicial} and \cite{Kadambari_2022_Distributed}. 
The sampling of graph edges using the line graph transformation is discussed in \cite{Kenta_edge_2022}. 
% Deep learning algorithms utilizing simplicial complexes to process data on the edges have been proposed by introducing nonlinearity to TSP.
% Graph neural network for edge data based on the Hodge Laplacian can be found in \cite{Roddenberry_2019}. 
% The work of \cite{Yang_2022} defined a convolutional neural network based on simplicial complexes. 
% The attention mechanism is merged with the Hodge Laplacian in \cite{Giusti_2022} and \cite{Lee_2022_SGAT}. 

The above TSP methods demonstrated good performance at processing signals on graph edges, but we may encounter additional challenges in realistic situations. 
Signals, in reality, often contain information about time dimension; when processing data on graph vertices that evolve with time, one could rely on graph adaptive filters, a combination of classical adaptive filters with graph shift operations. 
In GSP, the graph adaptive least mean squares (GLMS) algorithm was proposed for online estimation of vertex signals through $l_2$-norm optimization, assuming Gaussian noise and bandlimitedness of signal \cite{bib_LMS}.
Extensions of GLMS have improved its performance in various aspects. 
For instance, the normalized GLMS (GNLMS) incorporates spectral domain normalization in the bandlimited filter, enhancing convergence speed compared to GLMS \cite{bib_NLMS}. 
Other algorithms such as the Graph Least Mean $p^{th}$ (GLMP)  \cite{bib_LMP}, normalized GLMP \cite{yan_2022_NLMP}, and Graph Alternating Minimization \cite{Li_robust_2023} algorithms have increased robustness against impulsive noise.
The adaptive Graph-Sign (G-Sign) reduces computational complexity and increases robustness by employing an update function based on $l_1$ norm optimization \cite{yan_2022_sign, yan_2023_sign}. 
% Adaptive graph algorithms can be realized not only in the spectral domain but also in the Spatial domain; the diffusion GLMS was proposed as the spatial version of GLMS and the Graph-Sign-Diffusion was proposed as the spatial version of G-Sign \cite{yan_2023_sign}. 
% Kernel-based time-varying graph signal reconstruction method was also proposed using space-time models \cite{Romero_Kernel_2017}. 
Another class of algorithms that processes time-varying vertex signals combines time-series analysis techniques such as Vector Autoregressive (VAR), Vector Autoregressive Moving Average (VARMA), and generalized autoregressive conditionally heteroscedastic (GARCH) models GSP, resulting in the graph VAR \cite{Mei_2017}, graph VARMA \cite{isufi_2016_autoregressive, Isufi_Filtering_2017}, and graph GARCH \cite{Hong_GGARCH_2023}. 
Recently, the discussions of time-varying signals beyond the graph vertices have emerged: online water flow imputation on graph edges was attempted strictly using flow conservation in \cite{Money_2023_Online}; the simplicial VAR was proposed in \cite{Krishnan_2023_SVAR} on the same water flow imputation task.
The major drawback of implementing time-series analysis techniques on topological structures is that the multivariate nature of topological signals increases model complexity and poses challenges in acquiring accurate models. 
Research on the online estimation of signals on simplicial complexes is limited, particularly for the joint task of spatial signal imputation and temporal adaptive predictions.
 
Prior TSP and GSP methods have another limitation of primarily focusing on signals in a specific simplicial order.
In other words, the task to be solved is only on the vertices or the edges, but rarely both, not to mention that the majority of previous GSP and TSP methods are targeted for time-invariant scenarios. 
The need to process signals residing in multiple orders of simplices has drawn increased attention to the interactions and appearances of signals across these different simplicial orders.
In biological networks, transportation networks, and population mobility networks examples, the graph vertices and edges both contain signals, with the edge signals influencing the behavior of the vertex signals \cite{Kuruoglu_2016, yang_2021_Adaptive, Panagopoulos_2021}.
Understanding the dynamic interactions within these networks and signals from different simplicial orders is a crucial task for facilitating online tracking and predictive decision-making algorithms to enhance efficiency and reliability in real-world applications.
On a side note, data gathered from the real world is often noisy, meaning that the denoising of simplicial data should be considered when formulating simplicial representations.
These challenges comprise addressing topological irregularities, revealing data interactions across multiple dimensions, eliminating data noise, and capturing time-varying patterns. 
Therefore, it is a pressing necessity for an online algorithm that focuses on time-varying multi-order and multi-variate data on simplicial complexes to bridge the cross-space-time signal gaps. 
 
In this paper, we propose the Adaptive Joint Vertex-Edge Estimation (AJVEE) algorithm as a strategy aimed at addressing the previously mentioned focuses for time-varying signals existing on both the vertices and the edges by considering the dynamic interplay between the signals on the vertices and edges. 
To the best of our knowledge, this is the first paper to consider the joint online estimation of time-varying signals on both the vertices and the edges. 
The major contributions of this paper are listed as follows:
% \noindent{\bf{Contributions:}} 
\begin{itemize}
\item The AJVEE framework is proposed for adaptive estimation of time-varying vertex and edge signals on graphs by integrating two separate online estimations for vertices and edges and then into a single time-varying regression model. 
AJVEE utilizes a time-varying regression model to capture the interaction between vertex signals, with the intensity of these interactions represented by edge signals.
AJVEE can be further generalized to operate beyond just vertices and edges, allowing for more complex signal interactions.
\item An adaptive Least Mean Square procedure based on the Hodge Laplacian (ALMS-Hodge) is derived to support AJVEE. 
Unlike previous adaptive GSP algorithms that only address vertex estimation errors, ALMS-Hodge is generalized to higher-order structures such as edges and triangles.
ALMS-Hodge relies upon both aggregated signals and estimation errors to conduct missing signal imputation in the spatial domain and adaptive predictions in the temporal domain.
\end{itemize}

In Section~\ref{sec_background}, we will provide preliminary knowledge in representing graphs using simplicial complexes, as well as basic signal processing techniques on simplicial complexes. 
The ALMS-Hodge is introduced in Section~\ref{sec_ALMS}. 
Derivations and analysis of the AJVEE are included in Section~\ref{sec_joint}; we also provide a more generalized formulation of AJVEE applied to time-varying signals on simplicial complexes that are not restricted simply to the vertices and edges. 
Experiment results on both synthetic data and real data are discussed in Section~\ref{sec_results}. Section~\ref{sec_conclusion} concludes the paper with a brief discussion of potential future work. 
Table~\ref{notation} summarizes some frequent symbols seen in this paper. 

\begin{table}[hpbt]
\caption{Symbols and their meanings}
\begin{center}
    \begin{tabular}{|l|l|}
    \hline
    Symbol & Meaning \\
    \hline
    $k$ & Order of a simplex or simplicial complex\\
    $\mathcal{X}_k$ & $k$-simplex (vertex: $k$ = 0, edge: $k$ = 1)\\
    $\mathcal{K}$ & simplicial complex\\
    $\mathcal{G}$ & graph\\
    $t$ & time index \\
    $\boldsymbol{x}_k[t]\in\mathbb{R}^{N_k}$ & $k$-simplicial signal of time $t$\\
    $N_k$ & Cardinality of  $\boldsymbol{x}_k[t]$ \\
    $\mathbf{L}_k\in\mathbb{R}^{N_k\times N_k}$ & Hodge Laplacian matrix\\
    $\mathbf{D}_{k, \mathcal{S}}\in\mathbb{R}^{N_k\times N_k}$ & Sampling (Masking) Matrix\\
    $\mathbf{U}_k\in\mathbb{R}^{N_k\times N_k}$ & Eigenvector matrix of $\mathbf{L}_k$\\
    $\mathcal{F}_k$ & The frequency set of the bandlimited filter\\
    $\mathbf{U}_{k,\mathcal{F}}\in\mathbb{R}^{N_k\times |\mathcal{F}_k|}$ & Bandlimited eigenvector matrix of $\mathbf{L}_k$\\
    $\mathbf{B}_k\in\mathbb{R}^{N_{k-1}\times N_k}$ & Incidence matrix\\
    $\mathbf{I}$ & Identity matrix \\
    $\mathbf{\Sigma}_k\in\mathbb{R}^{N_k\times N_k}$ & Simplicial spectral filter \\
    $\hat{\boldsymbol{x}}_k[t]\in\mathbb{R}^{N_k}$ & Estimated $k$-simplicial signal of time $t$\\
    % $\hat{\boldsymbol{e}}[t]\in\mathbb{R}^{N_k}$ & Estimation error of time $t$\\
    $\hat{\boldsymbol{s}}_k[t]\in\mathbb{R}^{N}$ & SFT of $\hat{\boldsymbol{x}}_k[t]$\\
    % $\tilde{\boldsymbol{s}}_k[t]\in\mathbb{R}^{N}$ & SFT of $\tilde{\boldsymbol{x}}_k[t]$\\
    $\boldsymbol{\eta}_k[t]\in\mathbb{R}^{N_k}$ & Noise of the $k$-simplicial signal\\
    ${\boldsymbol{y}}_k[t]\in\mathbb{R}^{N_k}$ & Noisy observation of ${\boldsymbol{x}}_k[t]$\\
    $\mu_{k}$ & Update step size\\
    $r_k$ & Aggregation weight\\
\hline 
\end{tabular}
\label{notation}
\end{center}
\end{table}

\section{Preliminary knowledge}
\label{sec_background}
\subsection{Graph representation}
Let us begin with a vertex (node) set $\mathcal{V} = {\{v_1 ... v_{N_0}}\}$.
We will use the terms node and vertex interchangeably throughout the paper.
A $k$-simplex $\mathcal{X}_k$ is defined as a subset of $\mathcal{V}$ that has cardinality $k+1$. 
A simplicial complex $\mathcal{K}$ of order $K$ can be defined as a collection of $k$-simplices $\{\mathcal{X}_{k,1} ... \mathcal{X}_{k,N_k}\}$, with $k = 0 ... K$. 
A simplicial complex $\mathcal{K}$ has the property that for a simplex $\mathcal{X}_k \in \mathcal{K}$, if another simplex with a lower order $\mathcal{X}_{k-1}$ is a subset of $\mathcal{X}_k$, then $\mathcal{X}_{k-1} \in \mathcal{K}$ \cite{Barbarossa_2020}. 
Following this definition, a graph $\mathcal{G} = \{ \mathcal{V, E} \}$, can be represented as a simplicial complex, where the vertex set $\mathcal{V}$ is the 0-simplex-set and the edge set $\mathcal{E} = \{ e_1 ... e_{N_1} \}$ is connectivity between vertices and the 1-simplex-set. 
A subscript $k$ denotes which $k$-simplex each variable belongs to and distinguishes among different orders of simplex and simplicial complexes. 
The variable ${N_k}$ is used to represent the cardinality of the simplices in $\mathcal{K}$, where ${N_0} = |\mathcal{V}|$ is the number of vertices, ${N_1} = |\mathcal{E}|$ is the number of edges, and ${N_2}$ is the number of triangles. 

To represent the relationship between two simplices of order $k-1$ and $k$ in a simplicial complex, the incidence matrix $\mathbf{B}_k \in \mathbb{R}^{N_{k-1} \times N_{k}}$ is used where the rows of $\mathbf{B}_k$ correspond to the $N_k$ elements in $\mathcal{X}_{k-1}$ and the columns of $\mathbf{B}_k$ correspond to the $N_k$ elements in $\mathcal{X}_k$. 
The values assignment of $\mathbf{B}_k$ are as follows: if simplex $x_{k-1}(i)$ in $\mathcal{X}_{k-1}$ is adjacent to a simplex $x_k(j)$ in $\mathcal{X}_k$, then the $ij^{th}$ element of $\mathbf{B}_k$ will have magnitude $1$, with the sign determined by the predefined orientation in the $k$-simplex. 
We should emphasize that the orientation, defined by the ordering of the simplicial set, is not the same as the direction, especially when $k=1$ for the edges \cite{Schaub_2020_Random}.
The incidence matrix $\mathbf{B}_k$ is a boundary operator because $\mathbf{B}_k$ represents how $k-1$-simplices bounds their upper adjacent $k$-simplices. 
Similarly, we can define the dual of the boundary operator by the transpose of the incidence matrix $\mathbf{B}_{k}^T$, which is a coboundary operator \cite{Barbarossa_2020}.

\subsection{Signal Processing on graphs and simplicial complexes}
\label{sec_SFT_TSP}
A simplicial signal $\boldsymbol{x}_k = [x_{k,1} ... x_{k,N_k}]^T$ on a simplicial complex residing on the $k$-simplices is a mapping from the $k$-simplex to $\mathbb{R}^{N_k}$. 
In GSP, spectral operations on the vertex signal $\boldsymbol{x}_0$ are made possible by defining the Graph Fourier Transform (GFT) using the spectral decomposition of the graph matrix \cite{bib_LMS}. 
In TSP, the Hodge Laplacian matrix plays the role of the graph Laplacian matrix.
For a simplicial complex with order $K$, the Hodge Laplacians can be defined as follows: 
\begin{equation}
    \mathbf{L}_k = 
    \begin{cases}
    \mathbf{B}_1\mathbf{B}_1^T, & k = 0,\\
        \mathbf{B}_k^T\mathbf{B}_k+\mathbf{B}_{k+1}\mathbf{B}_{k+1}^T = \mathbf{L}_{k,l}+\mathbf{L}_{k,u},  
        & 0 < k < K,\\
            \mathbf{B}_{K}^T\mathbf{B}_{K}, &  k = K.
    \end{cases}
    \label{hodge_laplacian}
\end{equation}
The Hodge Laplacian $\mathbf{L}_k$ can be split into lower Hodge Laplacian $\mathbf{L}_{k,l} = \mathbf{B}_k^T\mathbf{B}_k$ and upper Hodge Laplacian $\mathbf{L}_{k,u} = \mathbf{B}_{k+1}\mathbf{B}_{k+1}^T$, contributing from the lower-adjacency and the upper-adjacency of $k$-simplices in a simplicial complex 
 \cite{Yang_2022_Simplicial}.

The TSP analogy of GFT is the Simplicial Fourier Transform (SFT) based on the spectral decomposition of the Hodge Laplacian matrix \cite{Barbarossa_2020}:
\begin{equation}
    \mathbf{L}_k = \mathbf{U}_k\mathbf{\Lambda}_k\mathbf{U}_k^T.
    \label{SFT}
\end{equation} 
The eigenvalue matrix is $\mathbf{\Lambda}_k = \text{diag}(\lambda_{k,1} ... \lambda_{k, N_k})$ with the eigenvalues $\lambda_{k, i}$ sorted in increasing order and considered as the frequencies. 
Smaller eigenvalues are considered as lower frequencies and larger eigenvalues are considered as higher frequencies.
Each eigenvalue in $\mathbf{\Lambda}_k$ has a corresponding eigenvector in the eigenvector matrix $\mathbf{U}_k = [\boldsymbol{u}_{k,1} ... \boldsymbol{u}_{k, N_k}]$. 
The forward SFT $\boldsymbol{s}_k = \mathbf{U}_k^T\boldsymbol{x}_k$ of a spatial domain signal $\boldsymbol{x}_k$ transforms it to the frequency domain signal $\boldsymbol{s}_k$.
The inverse SFT is $\boldsymbol{x}_k = \mathbf{U}_k\boldsymbol{s}_k$ transforms the frequency domain signal back to the spatial domain. 
It is worth mentioning that the GFT is a special case of the SFT for signals defined only on the graph vertices \cite{Barbarossa_2020}. 
The Hodge Laplacian is known as the graph Laplacian when $k=0$ and the graph Helmholtzian when $k=1$ \cite{Lim_2020_Hodge}. 
To process the signals residing on the edges, we can use SFT with $k=1$. 

With the SFT defined, we can apply predefined simplicial filters $\mathbf{\Sigma}$ to each frequency component in $\mathbf{L}_k$ to conduct spectral domain operations on simplicial complexes. 
A simple bandlimited filter in $\mathcal{X}_k$ based on a frequency set $\mathcal{F}_k$ is constructed by a diagonal matrix $\mathbf{\Sigma}_{\mathcal{F},k}$, where a $1$ on the $i^{th}$ diagonal element indicates that $\lambda_i \in \mathcal{F}_k$ and $0$ otherwise \cite{Barbarossa_2020}. 
We can define several spectral TSP filters and linearly combine them to obtain the desired frequency response. 
The following procedure is a simple yet complete bandlimited spectral filtering of $\boldsymbol{x}_k$:
\begin{equation}
    \boldsymbol{x}_k^\prime = \mathbf{U}_k\mathbf{\Sigma}_{\mathcal{F},k}\boldsymbol{s}_k =\mathbf{U}_k\mathbf{\Sigma}_{\mathcal{F},k}\mathbf{U}_k^T\boldsymbol{x}_k.
    \label{filter}
\end{equation}
If the signal $\boldsymbol{x}_k$ is bandlimited with frequencies $\mathcal{F}_k$, then $\boldsymbol{x}_k = \mathbf{U}_{k}\mathbf{\Sigma}_{\mathcal{F},k}\mathbf{U}_{k}^T\boldsymbol{x}_k$. 
To simplify notation and make use of the sparsity gained from eliminating some frequency components, we will set $\mathbf{U}_{k,\mathcal{F}} =$ support$(\mathbf{U}_k\mathbf{\Sigma}_{\mathcal{F},k}) $, where support() is the operation of dropping the column with all zero elements. 
Notice that the filter $\mathbf{\Sigma}_{\mathcal{F},k}$ is idempotent and self-adjoint, so with the simplified notation we have 
\begin{equation}
    \mathbf{U}_{k,\mathcal{F}}\mathbf{U}_{k,\mathcal{F}}^T = \mathbf{U}_k\mathbf{\Sigma}_{\mathcal{F},k}\mathbf{U}_k^T
\end{equation}
and 
\begin{equation}
    \mathbf{U}_{k,\mathcal{F}}^T\mathbf{U}_{k,\mathcal{F}} = \mathbf{I}.
\end{equation} 

The filtering operation in \eqref{filter} can be done using a spectral approach or a spatial approach \cite{Kadambari_2022_Distributed}.
To achieve similar effects as \eqref{filter} in the spatial domain, a predefined bandlimited filter is approximated using a series of shifted Chebyshev polynomials, and the spatial relationship is represented by the Hodge Laplacian. 
Assuming that spatial operations are conducted in the $k$-simplicies of simplicial complex $\mathcal{K}$, the shifted Chebyshev polynomial $T_p(\mathbf{L}_k)$ with order $P$ is defined as
\begin{equation}
    T_p(\mathbf{L}_k) = 
    \begin{cases}
    1, & \text{if } p = 0,\\
    \frac{2\mathbf{L}_k-\lambda_\text{max}}{\lambda_\text{max}}, & \text{if } p = 1,\\
    \frac{4\mathbf{L}_k-2\lambda_\text{max}}{\lambda_\text{max}}T_{p-1}(\mathbf{L}_k)-T_{p-2}(\mathbf{L}_k), & \text{if } p \geq 2.
    \label{eq_cheb}
    \end{cases}
\end{equation}
Using \eqref{eq_cheb}, the approximation of $\mathbf{U}_k\mathbf{\Sigma}_{\mathcal{F},k}\mathbf{U}_k^T$ in \eqref{eq_lp} is expressed as
\begin{equation}
\mathbf{U}_k\mathbf{\Sigma}_{\mathcal{F},k}\mathbf{U}_k^T \approx {\theta_0}+\sum_{p=1}^{P}\theta_p T_p(\mathbf{L}_k),
    \label{eq_cheb_approx}
\end{equation}
where $\theta_p$ is the coefficient of the Chebyshev polynomial. 

A special case of the bandlimited filter is the ideal low-pass filter, which is based on the (spatial) smoothness assumption of the signal \cite{Kenta_edge_2022}. 
The ideal low-pass filter $\mathbf{\Sigma}_{k,\mathcal{F}_{lp}}$ with cutoff frequency $\lambda_{k,c}$ can be defined using the frequency set $\mathcal{F}_{lp}$ and can be expressed as
\begin{equation}
    \mathbf{\Sigma}_{k,\mathcal{F}_{lp}} = \text{diag}(\boldsymbol{\sigma}_{k,\mathcal{F}_{lp}}),
    \label{eq_lp}
\end{equation}
where $\lambda_{k,i} \in \mathcal{F}_{lp} \; | \; i \leq c, \quad \lambda_{k,i} \notin \mathcal{F}_{lp} \; | \; i > c
$, and $1<c<N_k$ is the index of the cutoff frequency $\lambda_{k,c}$.
This means that the value assignment of the $i^{th}$ element of $\boldsymbol{\sigma}_{k,\mathcal{F}_{lp}}$ follows ${\sigma}_{k, \mathcal{F}_{lp}, i} = \mathbb{I}(i \leq c)$.

According to the TSP sampling theory for signals defined on $\mathcal{K}$, partial observation is modeled using a sampling operation on $\mathbf{D}_{k, \mathcal{S}}$ \cite{Barbarossa_2020}. 
The sampling set $\mathcal{S}_k$ defines the diagonal matrix masking operation, where the sampling matrix $\mathbf{D}_{k, \mathcal{S}} = \text{diag}(d_{k, 1} ... d_{k, N_k})$, where $d_{k,i} = 1$ if $x_{k,i} \in \mathcal{S}_k$ and $0$ otherwise \cite{Barbarossa_2020}. 
The sampling matrix $\mathbf{D}_{k, \mathcal{S}}$ is a diagonal matrix that is both idempotent and self-adjoint. 
When signal observations are missing due to various conditions or criteria other than by sampling, the matrix $\mathbf{D}_{k, \mathcal{S}}$ still denotes the observation mask.

Before proceeding to AJVEE, it is necessary to discuss how to reconstruct a static signal $\boldsymbol{x}_k$ when only partial observation of the signal is obtainable. 
With the assumption that $\boldsymbol{x}_k$ is a bandlimited signal with frequencies $\mathcal{F}_k$, a reconstruction of $\boldsymbol{x}_k$ is possible if the conditions $\|\mathbf{D}_{k, \bar{\mathcal{S}}}\mathbf{U}_{k,\mathcal{F}}\mathbf{U}_{k,\mathcal{F}}^T\|_2 = \|\mathbf{U}_{k,\mathcal{F}}\mathbf{U}_{k,\mathcal{F}}^T\mathbf{D}_{k, \bar{\mathcal{S}}}\|_2<1$ and $|\mathcal{F}_k|\leq|\mathcal{S}_k|$ are satisfied, where $\mathbf{D}_{k, \bar{\mathcal{S}}} = \mathbf{I} - \mathbf{D}_{k, \mathcal{S}}$ \cite{Barbarossa_2020}. 

\section{The ALMS-Hodge}
\label{sec_ALMS}
Before discussing the AJVEE framework, we would like to introduce a novel adaptive filtering algorithm named the ALMS-Hodge for online time-varying simplicial signal prediction on the $k$-simplices.  
Let us consider a graph $\mathcal{G}$ as a simplicial complex $\mathcal{K}$ containing time-varying signal $\boldsymbol{x}_k[t]$. 
In classical signal processing, a simple data model can be used to represent the time-varying dynamics of signals:
\begin{equation}
    \boldsymbol{x}_k[t+1] = \boldsymbol{x}_k[t]+\Delta_k[t],
    \label{eq_adaptive}
\end{equation}
where $\Delta_k[t]$ is the change that lead $\boldsymbol{x}_k[t]$ to $\boldsymbol{x}_k[t+1]$. 
Suppose we have another signal $\boldsymbol{y}_k[t]$, the noisy observation of $\boldsymbol{x}_k[t]$ that has missing elements. 
This signal $\boldsymbol{y}_k[t]$ is modeled using a sampling operation on $\mathbf{D}_{k, \mathcal{S}}$ and an i.i.d. additive Gaussian noise $\boldsymbol{\eta}_k[t]$ similar to what is found in GSP \cite{bib_LMS}:
\begin{equation}
    \boldsymbol{y}_k[t] = \mathbf{D}_{k, \mathcal{S}}(\boldsymbol{x}_k[t]+\boldsymbol{\eta}_k[t]).
\end{equation}
We could set up a cost function $\mathbf{J}(\hat{\boldsymbol{x}}_k[t])$ that minimizes the squared error between $\boldsymbol{y}_k[t]$ and $\boldsymbol{x}_k[t]$ due to the fact that an $l_2$-norm optimization problem yields the optimal solution under Gaussian noise assumption \cite{CHEN_2016_Variance}:  
\begin{equation}
        \mathbf{J}(\hat{\boldsymbol{x}}_k[t]) = f(\hat{\boldsymbol{x}}_k[t])+g(\hat{\boldsymbol{x}}_k[t]),
 \label{cost}
\end{equation}
where $f(\hat{\boldsymbol{x}}_k[t]) = \mathbb{E}\|\boldsymbol{y}_k[t]-{\mathbf{D}_{k, \mathcal{S}}}\mathbf{U}_{k,\mathcal{F}}\mathbf{U}_{k,\mathcal{F}}^T\hat{\boldsymbol{x}}_k[t]\|^2_2$ is the $l_2$ cost that we are aiming to optimize,  $\hat{\boldsymbol{x}}_k[t]$ is the estimation of the ground truth signal ${\boldsymbol{x}}_k[t]$, and $g(\hat{\boldsymbol{x}}_k[t])$ is a task-specific regularization in terms of simplicial aggregation that we will be deriving later. 
At $t=0$, the estimation $\boldsymbol{x}_k[0]$ is the initialization of the ALMS-Hodge. 
Here we assume that both the estimated signal $\hat{\boldsymbol{x}}_k[t]$ and  $\boldsymbol{x}_k[t]$ are bandlimited and share the same frequency component. 
This consistency is ensured because both $\boldsymbol{x}_k[t]$ and $\hat{\boldsymbol{x}}_k[t]$ use the same topology, meaning that the SFT and frequency components are defined by the same Hodge Laplacian. 
The filter design of $\mathbf{\Sigma}_{k, \mathcal{F}}$ can follow a universal filter such as the smoothness-based low-pass filter \cite{Kenta_edge_2022}.
Another option we can choose is to use the SFT to acquire the spectrum directly from the historical data or estimate the spectrum from the noisy observation.
We formulate \eqref{cost} as a convex least mean squares optimization problem similar to GSP on the vertices \cite{bib_LMS}:
\begin{equation}
    \min_{\hat{\boldsymbol{x}}_k[t]} \mathbf{J}(\hat{\boldsymbol{x}}_k[t]),
     \text{   s.t. } \mathbf{U}_{k,\mathcal{F}}\mathbf{U}_{k,\mathcal{F}}^T \hat{\boldsymbol{x}}_k[t] = \hat{\boldsymbol{x}}_k[t].
     \label{optimization}
\end{equation}
The solution of \eqref{optimization} could be found by stochastic gradient seen in classical and graph adaptive filtering \cite{bib_NLMS, bib_LMP, Diniz_2007_adaptive_filtering}.

In ALMS-Hodge, an essential step in online estimation is to effectively represent the change $\Delta_k[t]$ on each simplex using TSP techniques. 
Looking at \eqref{optimization}, we can represent $\Delta_k[t]$ as the change between two-time points, which is obtained by the (negative) gradient of $J(\hat{\boldsymbol{x}}_k[t])$. 
By letting $\Delta_k[t] = -\frac{\partial \mathbf{J}(\hat{\boldsymbol{x}}_k[t]) }{\partial \hat{\boldsymbol{x}}_k[t]}$ in \eqref{eq_adaptive}, an adaptive update function based on \eqref{eq_adaptive} for online simplicial signal estimation on order $k$ using the Hodge Laplacian can be derived as
\begin{equation}
        \hat{\boldsymbol{x}}_k[t+1] = \hat{\boldsymbol{x}}_k[t]+\Delta_k[t] = \hat{\boldsymbol{x}}_k[t] - \frac{\partial \mathbf{J}(\hat{\boldsymbol{x}}_k[t]) }{\partial \hat{\boldsymbol{x}}_k[t]}.
        \label{eq_update}
\end{equation}
Combining bandlimitedness of $\hat{\boldsymbol{x}}_k[t]$ and the idempotent property of $\mathbf{D}_{k, \mathcal{S}} $, the term $\boldsymbol{y}_k[t]-{\mathbf{D}_{k, \mathcal{S}}}\mathbf{U}_{k,\mathcal{F}}\mathbf{U}_{k,\mathcal{F}}^T\hat{\boldsymbol{x}}_k[t]$ can be written as $\mathbf{D}_{k, \mathcal{S}}(\boldsymbol{y}_k[t]-\mathbf{U}_{k,\mathcal{F}}\mathbf{U}_{k,\mathcal{F}}^T\mathbf{D}_{k, \mathcal{S}}\hat{\boldsymbol{x}}_k[t])$. 
The update function of our ALMS-Hodge is derived as
\begin{equation}
    \hat{\boldsymbol{x}}_k[t+1]  =  \hat{\boldsymbol{x}}_k[t]+\mu_k\mathbf{H}_k\boldsymbol{e}_k[t]+r_k\mathbf{R}_k\hat{\boldsymbol{x}}_k[t],
    \label{eq_update_2}
\end{equation}
where $-2\mathbf{H}_k{\boldsymbol{e}}_k[t] = \frac{\partial f(\hat{\boldsymbol{x}}_k[t])}{\partial \hat{\boldsymbol{x}}_k[t]}$ and  $\mathbf{R}_k\hat{\boldsymbol{x}}_k[t] = \frac{\partial g(\hat{\boldsymbol{x}}_k[t])}{\partial \hat{\boldsymbol{x}}_k[t]}$. 
The term $\boldsymbol{e}_k[t] = \mathbf{D}_{k, \mathcal{S}}(\boldsymbol{y}_k[t]-\hat{\boldsymbol{x}}_k[t])$ in \eqref{eq_update_2} is the estimation error of all sampled part of the signals.
Here, we can convert the spectral update into a general formulation of spatial update and use again the Chebyshev approximation in \eqref{eq_cheb_approx} to obtain a spatial update:
\begin{equation}
    \mathbf{H}_k =  \mathbf{U}_{k,\mathcal{F}}\mathbf{U}_{k,\mathcal{F}}^T\approx {\theta_0}+\sum_{p=1}^{P}\theta_p T_p(\mathbf{L}_{k}).
    \label{H_k}
\end{equation}
Step size parameters $\mu_k$ and $r_k$ are assigned to control the magnitude of the update terms as conventionally seen in classical adaptive filters \cite{Diniz_2007_adaptive_filtering}.
The frequencies are defined as larger eigenvalues corresponding to high frequencies, so we can define $\mathbf{\Sigma}_{k}$ as low-pass, high-pass, or band-limited filters similar to how these types of filters are designed in classical signal processing.
Viewed in the spectral domain, each update term $\mu_k\mathbf{H}_k\boldsymbol{e}_k[t]$ is applying the filter $\mathbf{\Sigma}_{k}$ to the SFT of the error component $\boldsymbol{e}_k[t]$.
As a result, at each time instance $t$, the term $\mathbf{H}_k\boldsymbol{e}_k[t]$ in \eqref{eq_update_2} ensures the update of the ALMS-Hodge is time-varying and is adaptively generated in the direction opposite to the error. 
It can be shown that the ALMS-Hodge in \eqref{eq_update_2} converges, for the details please check Appendix~\ref{app_convergence}. 

If the term $r_k\mathbf{R}_k\hat{\boldsymbol{x}}_k[t]$ is ignored as seen in previous adaptive GSP algorithms, the ALMS-Hodge will update solely based on the errors $\boldsymbol{e}_k[t]$  but not utilizing the signals $\boldsymbol{x}_k[t]$ itself.
Thus, it is advantageous if we define aggregation terms based on the underlying signal to enforce additional restrictions or enhancements. 
For example, if we want to emphasize the property $\mathbf{B}_k^T\mathbf{B}_k\hat{\boldsymbol{x}}_k[t] = 0$ on the lower adjacency of $\mathcal{X}_k$, then we can define an aggregation term $g(\hat{\boldsymbol{x}}_{k}[t]) = \|\mathbf{B}_{k}\hat{\boldsymbol{x}}_{k}[t]\|^2_2$ in the cost function \eqref{cost}. 
Following the optimization steps in \eqref{eq_update}, the resulting aggregation $\mathbf{R}_{k,l}\hat{\boldsymbol{x}}_k[t]$ in the update equation \eqref{eq_update_2} is
\begin{equation}
    r_{k}\mathbf{R}_{k,l}\hat{\boldsymbol{x}}_k[t] = \frac{\partial g(\hat{\boldsymbol{x}}_{k}[t])}{\partial \hat{\boldsymbol{x}}_{k}[t]} = 2\nu_k\mathbf{L}_{k,l}\hat{\boldsymbol{x}}_{k}[t],
    \label{eq_lower_adjacent_aggregation}
\end{equation}
where $ r_{k} = 2\nu_k$ is the weight for the aggregation. 
Similarly, we can define the upper adjacent aggregation $\mathbf{R}_{k,u}\hat{\boldsymbol{x}}_k[t]$ using the same approach by setting $h(\hat{\boldsymbol{x}}_{k}[t]) = \|\mathbf{B}_{k+1}^T\hat{\boldsymbol{x}}_{k}[t]\|^2_2$:
\begin{equation}
    \mathbf{R}_{k,u}\hat{\boldsymbol{x}}_k[t] = \mathbf{L}_{k,u}\hat{\boldsymbol{x}}_k[t].
        \label{eq_upper_adjacent_aggregation}
\end{equation}

Let us bring in the concept of Hodge decomposition.
A $\mathcal{X}_k$ simplicial signal can be decomposed as follows \cite{Barbarossa_2020}:
\begin{equation}
    \boldsymbol{x}_k[t] = \boldsymbol{x}_{k, H}[t]+\mathbf{B}_k^T\boldsymbol{x}_{k-1}[t]+\mathbf{B}_{k+1}\boldsymbol{x}_{k+1}[t].
    \label{eq_hodge_decomposition}
\end{equation}
The terms $\mathbf{B}_k^T\boldsymbol{x}_{k-1}[t]$ and $\mathbf{B}_{k+1}\boldsymbol{x}_{k+1}[t]$ are the signal components that can be induced from $\mathcal{X}_{k-1}$ and $\mathcal{X}_{k+1}$ signals respectively.  
The term $\boldsymbol{x}_{k, H}[t]$ in \eqref{eq_hodge_decomposition} is the harmonic component corresponding to the portion of the signal that cannot be induced from $\boldsymbol{x}_{k-1}$ or $\boldsymbol{x}_{k+1}$. 
It is worth mentioning that these three components are orthogonal \cite{Barbarossa_2020}. 
Remember in Section~\ref{sec_background}, the upper adjacency, lower adjacency, boundary, and coboundary of $\mathcal{X}_k$ can be defined using $\mathbf{B}_k$ and $\mathbf{B}_{k+1}$.
Inspired by the Hodge decomposition, at each time instance $t$, we can define regularization terms in terms of simplicial aggregation based on the boundary and coboundary. 
For example, if we want to aggregate the induced components from $\mathcal{X}_{k-1}$ onto $\mathcal{X}_{k}$, we can use the boundary aggregation
\begin{equation}
    \mathbf{R}_{k,b}\hat{\boldsymbol{x}}_k[t] = \mathbf{B}_{k}^T\hat{\boldsymbol{x}}_{k-1}[t].
        \label{eq_boundary_aggregation}
\end{equation}
Similarly, the coboundary aggregation can be defined as
\begin{equation}
    \mathbf{R}_{k,c}\hat{\boldsymbol{x}}_k[t] = \mathbf{B}_{k+1}\hat{\boldsymbol{x}}_{k+1}[t].
        \label{eq_coboundary_aggregation}
\end{equation}
Note that the above four aggregations can be linearly combined in AJVEE to achieve the desired effects. 

Paying attention to \eqref{cost} and \eqref{eq_update_2}, we see that the estimation error $\boldsymbol{e}_k[t]$ is only calculated using the sampled signals. 
To compensate for this in the aggregation, the aggregation parameters are split as $\mathbf{R}_k = \mathbf{R}_{k, \mathcal{S}}+\mathbf{R}_{k, \bar{\mathcal{S}}} = \mathbf{D}_{k, \mathcal{S}}\mathbf{R}_k+ \mathbf{D}_{k, \bar{\mathcal{S}}}\mathbf{R}_k$. 
Then, we assign different weightes to $\mathbf{R}_{k, \mathcal{S}}$ and $\mathbf{R}_{k, \bar{\mathcal{S}}}$:
\begin{equation}
    r_{k}\mathbf{R}_k\hat{\boldsymbol{x}}_k[t] = r_{k, \mathcal{S}}\mathbf{R}_{k, \mathcal{S}}\hat{\boldsymbol{x}}_k[t]+r_{k, \bar{\mathcal{S}}}\mathbf{R}_{k, \bar{\mathcal{S}}}\hat{\boldsymbol{x}}_k[t]. 
\end{equation}
The logic behind this setup is that different weights were assigned to the sampled and unsampled signals. 

Even though the aggregation $\mathbf{R}_k$ is task-specific and defined from the properties of the signal of interest, we can still gain some insights by looking at them in the spectral domain.
Generally speaking, by the Hodge decomposition \eqref{eq_hodge_decomposition}, the nonzero eigenvalue-eigenvector pairs in $\mathbf{B}_k\mathbf{B}_k^T$ are orthogonal to the nonzero eigenvalue-eigenvector pairs in $\mathbf{B}_{k+1}^T\mathbf{B}_{k+1}$.  
All the nonzero eigenvalue-eigenvector pairs in $\mathbf{L}_k$ are either from $\mathbf{B}_k\mathbf{B}_k^T$ or $\mathbf{B}_{k+1}^T\mathbf{B}_{k+1}$ \cite{Barbarossa_2020}. 
Another interesting property is $\mathbf{B}_{k}\mathbf{B}_{k+1} = 0$ because in the Hodge decomposition \eqref{eq_hodge_decomposition}, the harmonic component of a 
simplicial $\mathcal{X}_k$ signal cannot be induced from $\mathcal{X}_{k+1}$ or $\mathcal{X}_{k-1}$. 
From the orthogonality property, the harmonic components are associated with the zero eigenvalue-eigenvector pairs of $\mathbf{L}_k$ \cite{Barbarossa_2020}. 

There are three main advantages of the ALMS-Hodge. 
First, unlike the classical adaptive filter that learns the filters while making online estimation, our proposed ALMS-Hodge relies on assumptions such as spatial smoothness and bandlimitedness of the simplicial signal, then predefine universal TSP filters such as the low-pass filter.
This means that the ALMS-Hodge will not be restricted to a specific order of simplicial signals.
Additionally, not needing to learn the filters from the data reduces the complexity of the algorithm and simplifies the implementation in practice. 
Lastly, the ALMS-Hodge directly uses topological information of the data, which allows the missing simplicial signals to be estimated along with the temporal prediction.
This is a unique characteristic of GSP and TSP algorithms that is not seen in classical signal processing.

\begin{figure}[hp]
     \centering
     \begin{subfigure}{0.25\textwidth}
         \centering
         \includegraphics[trim={0 0 0 10pt },clip, width=\textwidth]{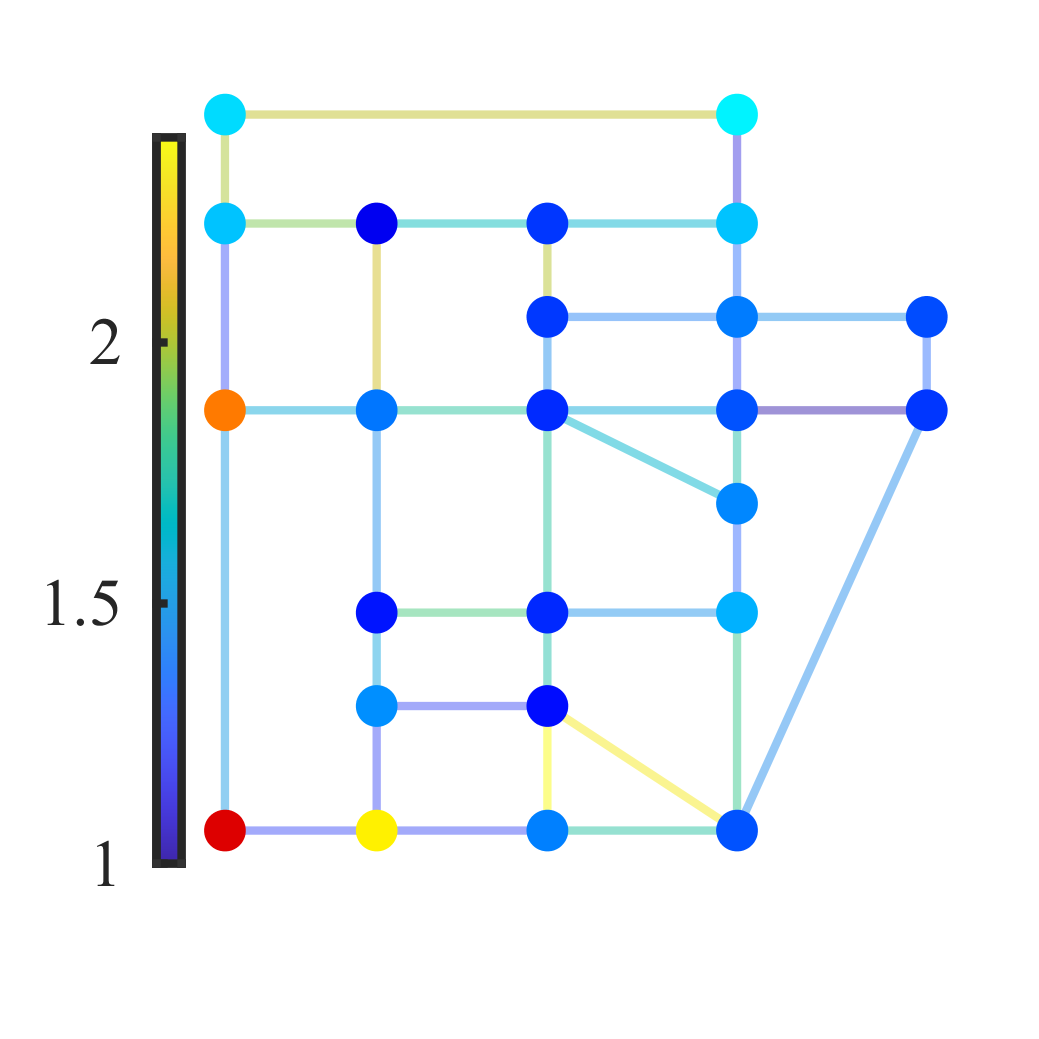}
                  \vspace{-30pt}
         \caption*{$t=1$}    
     \end{subfigure}
     \hspace{-17pt}
     \begin{subfigure}{0.25\textwidth}
         \centering
         \includegraphics[trim={0 0 0 10pt},clip, width=\textwidth]{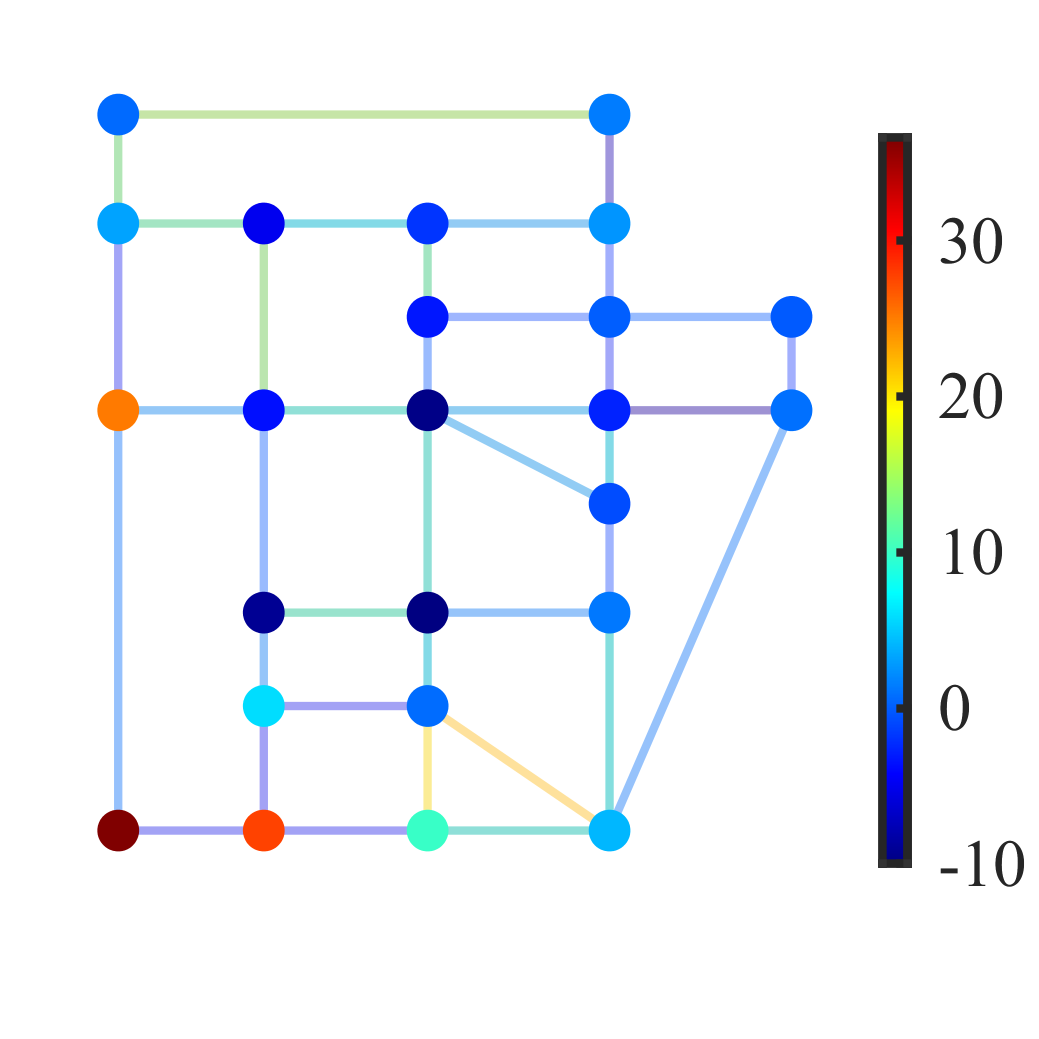}
                   \vspace{-30pt}
         \caption*{$t=180$}
     \end{subfigure}
        \caption{The Sioux Falls network with time-varying signals on both the vertices and the edges. (left color bar: edges, right color bar: vertices)}
        \label{fig1}
\end{figure}

\section{Adaptive joint vertex-edge estimation}
\label{sec_joint}
In this section, the AJVEE is proposed to target the case that the time-varying edge signal influences the time-varying vertex signal, from a simplicial diffusion perspective on a graph.
The ALMS-Hodge will be used to facilitate the operation of AJVEE.

\subsection{Vertex regression influenced by time-varying edge signals}
\label{sec_regression_data_model}
In AJVEE, a TSP approach is taken to represent data interactions occurring between two simplices of order $k=1$ and $k=0$, corresponding to the edges and the vertices respectively. 
A time-varying vertex regression with the regression parameters represented in the form of the time-varying edge signal is modeled in AJVEE instead of using fixed edge weights as seen in GSP. 
To begin, we will formulate a time-varying regression of the vertices by representing the regression parameters using edge time-varying signals: 
% \begin{equation}
%         \boldsymbol{x}_{0}[t+1] = \hat{\mathbf{L}}_{0}[t]\boldsymbol{x}_{0}[t] 
%     = \hat{\mathbf{B}}_{1}\hat{\mathbf{B}}_{1}^T\boldsymbol{x}_{0}[t],
%     \label{regression}
% \end{equation}
\begin{equation}
        \boldsymbol{x}_{0}[t+1] = \hat{\mathbf{L}}_{0}[t]\boldsymbol{x}_{0}[t] 
    = {\mathbf{B}}_{1}\text{diag}(|\boldsymbol{x}_1[t]|){\mathbf{B}}_{1}^T\boldsymbol{x}_{0}[t],
    \label{regression}
\end{equation}
where $\hat{\mathbf{L}}_{0}[t]$ is a time-varying regression matrix derived from a graph topology containing a temporal edge signal, and $t$ is the time index.
% The term $\hat{\mathbf{B}}_{1}$ in equation~\eqref{regression} can be expressed as 
% \begin{equation}
%         \hat{\mathbf{B}}_{1} = (\mathbf{B}_{1}\text{diag}(\text{sign}(\boldsymbol{x}_1[t])\circ|\boldsymbol{x}_1[t]|^{0.5})),
%         \label{eq_reg_weights}
% \end{equation}
% where $\boldsymbol{x}_1[t]$ is the time-varying edge signal, diag$(\boldsymbol{x}_1[t])$ forms a diagonal matrix from $\boldsymbol{x}_1[t]$, $|\boldsymbol{x}_1[t]|^{0.5}$ is the element-wise power of the absolute value of $\boldsymbol{x}_1[t]$, and sign() computes the element-wise sign. 
% \begin{equation}
%         \hat{\mathbf{B}}_{1} = (\mathbf{B}_{1}\text{diag}(\text{sign}(\boldsymbol{x}_1[t])\circ|\boldsymbol{x}_1[t]|^{0.5})),
%         \label{eq_reg_weights}
% \end{equation}
We should emphasize that this process is fundamentally different from spatial node aggregation using purely the graph Laplacian $\mathbf{L}_0$. 
The regression matrix $\hat{\mathbf{L}}_{0}[t]$ is formed by mapping time-varying edge signals $\boldsymbol{x}_1[t]$ onto the topology and should not simply be perceived as a static weighted version of the graph Laplacian matrix ${\mathbf{L}}_{0}$.
By combining $\hat{\mathbf{L}}_{0}[t]$ time-varying edge signals, we result in a time-varying influence of $\boldsymbol{x}_1[t]$ on $\boldsymbol{x}_{0}[t]$ in the form of time-varying regression matrix $\hat{\mathbf{L}}_{0}[t]$ that models the change of vertex signal $\boldsymbol{x}_{0}[t]$. 

A diffusion process on a graph is when vertex signals propagate through the adjacency relationships to their locally connected neighbors \cite{Bick_2023_higher_order, Xie_2022}. 
If we break the matrix multiplication in \eqref{regression} down, we will see that 
\begin{equation}
    {x}_i[t+1] = \sum^{N}_{j=1}\hat{{l}}_{ij}x_i[t],
    \label{eq_relation_2}
\end{equation}
where $\hat{{l}}_{ij}$ is the $i^{th}$ row and $j^{th}$ column of $\hat{\mathbf{L}}_0[t]$, and  $x_i[t]$ is the $i^{th}$ element of $\boldsymbol{x}_0[t]$. 
Realizing that in $\hat{\mathbf{L}}_0[t]$, $\hat{{l}}_{ij}$ is non-zero only when there is an edge, each summation in equation~\eqref{eq_relation_2} essentially is calculating the vertex signal at $x_i$ minus all the neighboring vertex signals from 1-hop away. 
When repeating \eqref{eq_relation_2} for all $N_0$ vertices, each vertex will be aggregated to its 1-hop adjacent neighbor. 
If we look at \eqref{eq_relation_2} in a diffusion aspect, it essentially means that all the vertices will be propagated to their connected neighbors, meaning that the aggregation done in \eqref{eq_relation_2} is equivalent to a diffusion. 

Following the above logic, the $p^{th}$ power of $\hat{\mathbf{L}}_{0}$ will be a time-varying diffusion in the $p$-hop neighborhood.
Summing all the terms for $p = 0...P$ gives us a combined $1$-hop to $P$-hop regression:
\begin{equation}
    \hat{\mathbf{H}}_{0}[t] = \sum_{p=0}^{P}\hat{\theta}_p \hat{\mathbf{L}}_{0}^{p}[t],
    \label{eq_cheb_H_2}
\end{equation}
where $\hat{\theta}_p$ is introduced as polynomial coefficients. 
The diffusive nature of this formulation means that missing data on the vertices can be imputed from neighborhood connections as the diffusion progresses.
Moreover, when the weights $\hat{\theta}_p$ in \eqref{eq_cheb_H_2} are defined based on a properly designed simplicial filter, simplicial denoising can be achieved with this formulation; a spectral analysis will be conducted in the next subsection as a guide of how to define the weights. 
The expression in \eqref{eq_cheb_H_2} can be realized by the Chebyshev approximation shown in \eqref{eq_cheb_approx}.
If we plug $\hat{\mathbf{L}}_{0}[t]$ into \eqref{eq_cheb_approx} and rearrange the terms, the resulting expression becomes a spatial diffusion represented by $\hat{\mathbf{L}}_{0}[t]$: 
\begin{equation}
    {\theta_0}+\sum_{p=1}^{P}\theta_p T_p(\hat{\mathbf{L}}_{0}[t]) = \sum_{p=0}^{P}\hat{\theta}_p \hat{\mathbf{L}}_{0}^p[t] = \hat{\mathbf{H}}_{0}[t],
    \label{eq_cheb_var}
\end{equation}
where now $\hat{\theta}_p$ is the polynomial coefficient instead of $\theta_p$. 

Another approach to forming the time-varying vertex regression in \eqref{regression} is to use spectral filters by setting the SFT in \eqref{SFT} to $k=0$ to form a simplicial (node) filter $\mathbf{\Sigma}_{0, \mathcal{F}}$ to approximate $\hat{\mathbf{L}}_{0}[t]$.
Let us represent the filter $\mathbf{\Sigma}_{0, \mathcal{F}}$ using a function $h(\mathbf{\Lambda}_{0}[t]) = \mathbf{\Sigma}_{0, \mathcal{F}}$. 
Then, the regression \eqref{regression} in terms of the spectral filtering in \eqref{filter} is
\begin{equation}
    \hat{\mathbf{U}}_{0}[t] h(\mathbf{\Lambda}_{0}[t]) \hat{\mathbf{U}}_{0}^T[t]\boldsymbol{x}_{0}[t] = h(\hat{\mathbf{L}}_{0}[t])\boldsymbol{x}_{0}[t]. 
    \label{H_and_h}
\end{equation}
Applying the logic of using a shifted Chebyshev polynomial $T_p(\mathbf{L}_{0}[t])$ similar to what is shown in \eqref{eq_cheb}, we can approximate \eqref{H_and_h} with the $p$-hop time-varying regression in \eqref{eq_cheb_H_2}:
\begin{equation}
         h(\hat{\mathbf{L}}_{0}) \approx  {\theta_0}+\sum_{p=1}^{P}\theta_p T_p(\hat{\mathbf{L}}_{0}[t]) = \sum_{p=0}^{P}\hat{\theta}_p \hat{\mathbf{L}}_{0}^{p}[t] = \hat{\mathbf{H}}_{0}[t]. 
    \label{eq_cheb_H_k}
\end{equation}
Equation \eqref{eq_cheb_H_k} establishes the relationship between the spectral domain and the spatial domain, illustrating the underlying filter $\mathbf{\Sigma}_{0, \mathcal{F}}$ can be applied from two different perspectives. 

There is a limitation of using spectral methods in practice.  
The simplicial filter $\mathbf{\Sigma}_{0, \mathcal{F}}$ is defined by the frequency set $\mathcal{F}_{0}$ operating on the frequencies (eigenvalues) of $\hat{\mathbf{L}}_{0}[t]$. 
This means that spectral filtering operation in \eqref{H_and_h} requires the eigendecomposition at every time instance when the signals $\boldsymbol{x}_1[t]$ changes due to the time-varying formulation in \eqref{regression}. 
Moreover, the eigendecomposition of the graph Laplacian suffers from high computational cost and numerical instabilities when the graph topology is large \cite{Shuman_2011_Chebyshev}. 
The same limitation can be found in TSP for the Hodge-Laplacian \cite{Kadambari_2022_Distributed}. 
Thus, the spatial formulations are preferred in practice while the spectral formulations provide a signal processing analytical perspective. 

For simplicity, we assume that the topology is known, but the signals on the vertices and edges are noisy or have missing entries at the current time $t$.
Preferably, the topology does not change over time. 
An example of synthetic data based on a real-world traffic network of the above data model is shown in Fig.~\ref{fig1}, where time-varying signals are generated on both the graph vertices and the edges. 
Even though the formulations in \eqref{eq_cheb_H_2} provide diffusion-based vertex regressions under the influence of edge signals, they consider only the diffusion perspective but not the online estimation perspective. 
We will improve this formulation by considering the temporal changes to achieve adaptive estimation.

\subsection{Joint estimation of vertex and edge signals}
AJVEE is targeted at cases where time-varying edge signals influence time-varying vertex signals. 
% For simplicity, we will assume the $\mathcal{X}_k$ signals are time-varying but the dynamics are modeled by time-invariant aggregation $\hat{\mathbf{H}}_k$ instead of the time-varying regression seen in the $\mathcal{X}_{k-1}$ signals.
We can specify the ALMS-Hodge onto the graph edges by specifying $k = 1$, resulting in $\mathbf{H}_1 = \sum_{p=0}^{P}\hat{\theta}_{1,p} \mathbf{L}_1^p$.
Using Chebyshev approximation \ref{eq_cheb_approx}, the AJVEE update based on the ALMS-Hodge on the edge signals can then be obtained as:
\begin{equation}
\begin{split}
        \hat{\boldsymbol{x}}_1[t+1]  =  \hat{\boldsymbol{x}}_1[t]+\mu_1\sum_{p=0}^{P}\hat{\theta}_{1,p} \mathbf{L}_1^p\boldsymbol{e}_1[t]
        \\
        +r_{1, \mathcal{S}}\mathbf{R}_{1, \mathcal{S}}\hat{\boldsymbol{x}}_1[t]+r_{1, \bar{\mathcal{S}}}\mathbf{R}_{1, \bar{\mathcal{S}}}\hat{\boldsymbol{x}}_1[t].
\end{split}
    \label{eq_AJVEE_k}
\end{equation}

To define the AJVEE expression on the graph vertices, we specify $k=0$ in the ALMS-Hodge. 
However, AJVEE will be following our time-varying regression data model defined in Section~\ref{sec_regression_data_model} on the graph vertices. 
The vertex signal $\boldsymbol{x}_{0}[t]$  is under the influence of the time-varying edge signal $\boldsymbol{x}_{1}[t]$ so the regression parameters are the current step edge signal estimate $\hat{\boldsymbol{x}}_1[t]$ obtained at the previous time step $t-1$.
As a result, the matrix $\mathbf{H}_k$ in ALMS-Hodge in \eqref{eq_update_2} is replaced with the time-varying vertex regression $\hat{\mathbf{H}}_{0}[t] =\sum_{p=0}^{P}\hat{\theta}_{0,p} (\hat{\mathbf{L}}_{0}[t])^p$ to incorporate the influence of the time-varying edge signal $\boldsymbol{x}_1[t]$ onto the time-varying vertex signals $\boldsymbol{x}_0[t]$.
The vertex update strategy in AJVEE based on ALMS-Hodge is
\begin{equation}
\begin{split}
    \hat{\boldsymbol{x}}_{0}[t+1]  = \hat{\boldsymbol{x}}_{0}[t]+\mu_{0}\sum_{p=0}^{P}\hat{\theta}_{0,p} (\hat{\mathbf{L}}_{0}[t])^p\boldsymbol{e}_{0}[t]
    \\
    +r_{0, \mathcal{S}}\mathbf{R}_{0, \mathcal{S}}\hat{\boldsymbol{x}}_0[t]+r_{0, \bar{\mathcal{S}}}\mathbf{R}_{0, \bar{\mathcal{S}}}\hat{\boldsymbol{x}}_0[t], \label{eq_AJVEE_k-1}
\end{split}
\end{equation} 
where  $\hat{\boldsymbol{x}}_{0}[t]$ is the estimation of vertex signals at time $t$, and the residual $\boldsymbol{e}_{0}[t] = \boldsymbol{y}_{0}[t] - \boldsymbol{x}_{0}[t]$ is the estimation error. 

Instead of separately estimating the vertex and edge signals, a joint estimation can be done by the simultaneous operation of \eqref{eq_AJVEE_k} and \eqref{eq_AJVEE_k-1}.
We define a joint operator $\mathbf{H}_J[t]$ to incorporate the adjacent relationship within one simplex order together with the incidence relationship between different orders of simplicity.
Assuming that the signals are on the vertices (0-simplex) and edges (1-simplex), and also assuming that we have the triangle (2-simplex) adjacency but not the 2-simplicial signals, we can formulate such joint shift operator on the error terms $\boldsymbol{e}_0[t]$ and $\boldsymbol{e}_1[t]$ by
\begin{equation}
    \mathbf{H}_J[t] = \text{blkdiag}(\mu_{0}\hat{\mathbf{H}}_0[t], \mu_{1}\mathbf{H}_1),
    \label{eq_H_J}
\end{equation}
where blkdiag$()$ is the block-diagonal concatenation of two matrices. 
Similarly, the signal aggregation matrices are concatenated as
\begin{equation}
\begin{split}
    & \mathbf{R}_{J,\mathcal{S}} = \text{blkdiag}(r_{0, \mathcal{S}}\mathbf{R}_{0,\mathcal{S}}, r_{1, \mathcal{S}}\mathbf{R}_{1,\mathcal{S}}) \text{ and } \\
    & \mathbf{R}_{J,\bar{\mathcal{S}}} = \text{blkdiag}(r_{0, \bar{\mathcal{S}}}\mathbf{R}_{0,\bar{\mathcal{S}}}, r_{1, \bar{\mathcal{S}}}\mathbf{R}_{1,\bar{\mathcal{S}}}).
    \label{eq_R_J}
\end{split}
\end{equation}
Accordingly, the signals, observations, and the estimation error should also be concatenated:
\begin{equation}
\begin{split}    
&\hat{\boldsymbol{x}}_J[t] = \text{vec}(\hat{\boldsymbol{x}}_0[t], \hat{\boldsymbol{x}}_1[t]), \\
&{\boldsymbol{y}}_J[t] = \text{vec}({\boldsymbol{y}}_0[t], {\boldsymbol{y}}_1[t]),
\text{ and } \\
    &{\boldsymbol{e}}_J[t] = \text{vec}({\boldsymbol{e}}_0[t], {\boldsymbol{e}}_1[t]).
\end{split}
\end{equation}
The update function of the AJVEE can be constructed as
\begin{equation}
            \hat{\boldsymbol{x}}_{J}[t+1]  = \hat{\boldsymbol{x}}_{J}[t]+\mathbf{H}_J[T]\boldsymbol{e}_{J}[t]
        +\mathbf{R}_{J, \mathcal{S}}\hat{\boldsymbol{x}}_J[t]+\mathbf{R}_{J, \bar{\mathcal{S}}}\hat{\boldsymbol{x}}_J[t]. \label{eq_update_joint}
\end{equation}
The individual vertex signal and edge signal can be obtained by mapping the corresponding elements in $\boldsymbol{x}_J[t+1]$ back into $\boldsymbol{x}_0[t+1]$ and $\boldsymbol{x}_1[t+1]$.

The AJVEE operates in an online fashion where the time-varying edge signal and the time-varying vertex signal are estimated jointly using \eqref{eq_update_joint} at each time instance. 
AJVEE simultaneously performs missing signal imputation as it makes predictions because of the incorporation of simplicial diffusion. 
Additionally, from the spectral filter aspect, the simplicial filter within AJVEE ensures denoising is done along the update. 
These advantages and characteristics of AJVEE enable AJVEE to overcome the challenges of representing topological irregularity, estimating unknown data from known data, removing noise present in the data, and extracting cross-space-time variations we pointed out at the beginning of Section~\ref{sec_intro}.
% It should be emphasized that the AJVEE discussed here is not simply estimating edge weights from graph vertex signals nor a link prediction task; the graph edges have the additional characteristics of containing signals on the 1-simplex.
The complete AJVEE algorithm is summarized in Algorithm~\ref{algorithm_joint_node_edge}. 
For the spectral analysis of the AJVEE from the TSP perspective, please refer to Appendix~\ref{app_spectral_analysis_AJVEE}.

\begin{algorithm}[htbp]
\caption{Adaptive joint vertex-edge estimation}
\begin{algorithmic}
    \State Given the (unweighted) graph topology $\mathcal{G}[t]|_{t=0}$, spectral filters $\mathbf{\Sigma}_0$ and $\mathbf{\Sigma}_{1}$     
    \State Initialize $\hat{\boldsymbol{x}}_0[0]$ and $\hat{\boldsymbol{x}}_1[0]$
    \State Construct $\mathbf{H}_1$ and $\hat{\mathbf{H}}_0[t]$
    \State Construct $\mathbf{R}_1$ and $\mathbf{R}_0$ based on prior knowledge using \eqref{eq_lower_adjacent_aggregation},  \eqref{eq_upper_adjacent_aggregation}, \eqref{eq_boundary_aggregation}, or \eqref{eq_coboundary_aggregation}
    \State Construct $\mathbf{H}_J[t]$ \eqref{eq_H_J} and $\mathbf{R}_J[t]$ using \eqref{eq_R_J}
    \While{there are new observations $\boldsymbol{y}_0[t]$ and $\boldsymbol{y}_1[t]$}
        \State Define the changed components in $\hat{\mathbf{H}}_J[t]$ and $\mathbf{R}_J$ again 
        \State Output joint signal $\hat{\boldsymbol{x}}_J[t+1]$ from $\boldsymbol{y}_J[t]$ using \eqref{eq_update_joint}
    \EndWhile
\end{algorithmic}
\label{algorithm_joint_node_edge}
\end{algorithm}

\subsection{Generalizing AJVEE to high-ordered simplicial complexes}
\label{sec_graph_AJVEE}
In this section, the AJVEE is defined given a more generalized scenario where the time-varying simplicial signals on $\mathcal{X}_k$ are influenced by the time-varying signals simplicial signals on $\mathcal{X}_{k+1}$. 
Given the topology, we can rely on defining the ALMS-Hodge on  $\mathcal{X}_k$ and $\mathcal{X}_{k+1}$ shown in Section~\ref{sec_ALMS}.
First, the $k$-simplex regression in \eqref{regression} will need to have the parameters defined using the $k+1$-simplex.
Then, we can define the joint terms 
\begin{equation}
\begin{split}    
&\hat{\boldsymbol{x}}_J[t] = \text{vec}(\hat{\boldsymbol{x}}_k[t], \hat{\boldsymbol{x}}_{k+1}[t]), \\
&{\boldsymbol{y}}_J[t] = \text{vec}({\boldsymbol{y}}_k[t], {\boldsymbol{y}}_{k+1}[t]),
\text{ and } \\
    &{\boldsymbol{e}}_J[t] = \text{vec}({\boldsymbol{e}}_k[t], {\boldsymbol{e}}_{k+1}[t]),
\end{split}
\end{equation}
where $\hat{\boldsymbol{x}}_J[t]$ is the joint time-varying $\mathcal{X}_k$ and $\mathcal{X}_{k+1}$ signals, ${\boldsymbol{y}}_J[t]$ is the joint observations, and ${\boldsymbol{e}}_J[t]$ is the joint estimation error. 
To formulate the proper AJVEE on simplicial complexes that are higher-ordered compared with graph vertices and edges, we will have matrices defined using the TSP techniques discussed in Section~\ref{sec_background}:
\begin{equation}
\begin{split}
    &\mathbf{H}_J[t] = \text{blkdiag}(\mu_{k}\hat{\mathbf{H}}_k[t], \mu_{k+1}\mathbf{H}_{k+1}), \\
    & \mathbf{R}_{J,\mathcal{S}} = \text{blkdiag}(r_{k, \mathcal{S}}\mathbf{R}_{k,\mathcal{S}}, r_{{k+1}, \mathcal{S}}\mathbf{R}_{{k+1},\mathcal{S}}) \text{ and } \\
    & \mathbf{R}_{J,\bar{\mathcal{S}}} = \text{blkdiag}(r_{k, \bar{\mathcal{S}}}\mathbf{R}_{k,\bar{\mathcal{S}}}, r_{{k+1}, \bar{\mathcal{S}}}\mathbf{R}_{{k+1},\bar{\mathcal{S}}}).
    \label{eq_R_J_k}
\end{split}
\end{equation}
The update scheme of AJVEE still follows \eqref{eq_update_joint} after plugging the above matrices into \eqref{eq_update_joint}.
However, conceptually, because we are defining all operations using the Hodge Laplacian and TSP, for $k \neq 0$, AJVEE is no longer defined on the vertices and edges.
For example, if $k = 2$, then AJVEE becomes an algorithm that jointly estimates the time-varying signals on the triangles (2-simplex) that are under the influence of the time-varying signals on the tetrahedrons (3-simplex).
The general case of AJVEE is illustrated in Algorithm~\ref{algorithm_AJVEE_general}.

\begin{algorithm}
\caption{Joint estimation of signals in $\mathcal{X}_{k}$ and $\mathcal{X}_{k+1}$}
\begin{algorithmic}
    \State Given: (unweighted) graph topology $\mathcal{G}[t]|_{t=0}$, spectral filters $\mathbf{\Sigma}_k$ and $\mathbf{\Sigma}_{k+1}$     
    \State Initialize $\hat{\boldsymbol{x}}_k[0]$ and $\hat{\boldsymbol{x}}_{k+1}[0]$
    \State Construct $\mathbf{H}_k$ and $\hat{\mathbf{H}}_{k+1}[t]$
    \State Construct $\mathbf{R}_k$ and $\mathbf{R}_{k+1}$ based on prior knowledge using \eqref{eq_lower_adjacent_aggregation},  \eqref{eq_upper_adjacent_aggregation}, \eqref{eq_boundary_aggregation}, or \eqref{eq_coboundary_aggregation}
    \State Construct $\mathbf{H}_J[t]$ and $\mathbf{R}_J[t]$ using \eqref{eq_R_J_k}
    \While{there are new observations $\boldsymbol{y}_k[t]$ and $\boldsymbol{y}_{k+1}[t]$}
        \State Define the changed components in $\hat{\mathbf{H}}_J[t]$ and $\mathbf{R}_J$ again 
        \State Output joint signal $\hat{\boldsymbol{x}}_J[t+1]$ from $\boldsymbol{y}_J[t]$ using \eqref{eq_update_joint}
    \EndWhile
\end{algorithmic}
\label{algorithm_AJVEE_general}
\end{algorithm}

\section{Experiment Results and Discussion}
\label{sec_results}
\subsection{Experiment settings}
The datasets we conduct the experiments on are the Sioux Falls transportation network, the Anaheim transportation network, and the England COVID-19 dataset. 
All the experiments are conducted in MATLAB 2022a and are repeated $N = 100$ times unless specified otherwise. 
In the experiment, all the tested algorithms will use lowpass filters defined in \eqref{eq_lp}.
The diffusion initialization strategy proposed in \cite{yan_2023_sign} is used to initialize the signals for all the experiments. 
All algorithms running on the same task will have the same filters with the same passband to ensure the fairness of comparison. 
The performance of all the experiments is measured using the normalized mean squared error (NMSE):
% \begin{equation}
%     \text{NMSE}_k[t]=\frac{1}{N} \sum_{i=1}^{N}\frac{(\hat{\boldsymbol{x}}_1[t]-\boldsymbol{x_1}[t])^2_i}{\|(\hat{\boldsymbol{x}}_1[t]-\boldsymbol{x_1}[t])_i\|^2_2},
%     \label{NMSE}
% \end{equation}
\begin{equation}
    \text{NMSE}_k[t]=\frac{1}{N} \sum_{i=1}^{N}\frac{\|(\hat{\boldsymbol{x}}_k[t]_i-\boldsymbol{x}_k[t])\|^2_2}{\|(\boldsymbol{x}_k[t])\|^2_2},
    \label{NMSE}
\end{equation}
where $\|(\hat{\boldsymbol{x}}_k[t]_i-\boldsymbol{x}_k[t])\|^2_2$ is the squared error on the $i^{th}$ run. 

The topology for the Sioux Falls transportation network and the Anaheim transportation network are gathered from the real world \cite{transportation}. 
The Sioux Falls network consists of $N_0 = 24$ vertices and $N_1 = 38$ edges and the Anaheim network has $N_0 = 406$ vertices and $N_1 = 624$ edges. 
These two transportation networks have provided static traffic flows on the edges for the traffic under equilibrium \cite{transportation}.
The vertex and edge signal missing rates for the Sioux Falls network is  $26\%$ and for the Anaheim network is $30\%$. 
The missing signal observations in the Sioux Falls network are modeled using the spectral sampling strategy discussed in \cite{bib_NLMS} and the missing signal observations in the Anaheim network are set to be random for each experiment run but will remain fixed within an experiment run. 
The edge signals are generated synthetically by combining several sinusoidal functions to make the static signal provided by the dataset time-varying:  
\begin{equation}
    \boldsymbol{x}_1[t] = \sum^P_{p=1}\boldsymbol{a}_pf_p[t],
    \label{synthetic_edge}
\end{equation}
where the signals $\boldsymbol{a}_p$ are static edge signals and the term $f_p[t]$ is a sinusoidal function.  
The ground truth vertex signals are synthetically generated with additive i.i.d. Gaussian noise $\epsilon_{0}[t]$ with zero mean and standard deviation of $=0.2$ on both the Sioux Falls network and the Anaheim network:
\begin{equation}
    \boldsymbol{x}_{0}[t+1] = \hat{\mathbf{H}}_{0}[t]\boldsymbol{x}_{0}[t]+\epsilon_{0}[t].
    \label{eq_RegressionSignalModle_1}
\end{equation}
We should point out that even though the vertex signal is synthetic, which seems a bit abstract, it is built using properties extracted from the underlying real-world topology. 
In the context of a transportation network, one may perceive the vertex signal as a measurement of congestion at an intersection or a number of the crowd at a place of interest. 
The magnitude of the vertex signal will be proportional to the amount of traffic on the edges. 
Using AJVEE, we aim to predict the congestion or population on the vertices using the predicted traffic on the edges. 
Since this paper is not focused on transportation network models, we will not discuss the meaning of vertex signals further and will use them purely as the target signal to be estimated.
Figure.~\ref{fig1} illustrates an example of such a time-varying vertex signal under the influence of time-varying edge signals on the Sioux Falls network. 

The England COVID-19 dataset consists of a series of 61 dynamic graphs \cite{Panagopoulos_2021}. 
In the experiment, we perceive every newly appeared edge at the topology in time $t$ compared to the previous topology at $t-1$ as an edge with signals missing, meaning that the corresponding diagonal entry in $\mathbf{D}_{1,\mathcal{S}}$ is $0$. 
All 61 graphs have $N_0 = 129$ vertices, with a different number of edges $N_1[t]$ on each day. 
The vertices represent different regions in England, recording the number of COVID-19 cases in the region each day, and the edges represent population mobility between regions; self-loops represent the mobility within the region. 
The upper row of Figure~\ref{Covid_figs} displays the England COVID-19 dataset at three different time points after our preprocessing. 
Even though there is no data on the $2$-simplex in any of the datasets, the upper adjacency of the edges can still be represented in $\mathbf{L}_{1,u}$ because the topological structures still exist. 

The AJVEE proposed in Algorithm~\ref{algorithm_joint_node_edge} will be examined on the ability to jointly reconstruct time-varying vertex and edge signals under the conditions that the edge signal influences the vertex signal.
Since ALMS-Hodge is the building block of AJVEE, we will conduct preliminary experiments on the ALMS-Hodge for both steady-state and time-varying simplicial signals in the $k$-simplices prior to the AJVEE experiments. 

\begin{figure}[htbp]
     \centering
     \begin{subfigure}{0.48\textwidth}
         \centering
         \includegraphics[trim={180 345 180 350},clip,width=\textwidth]{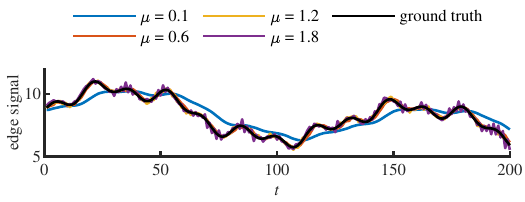}
         \caption{The online estimation of one selected edge.}
         \label{fig_change_mu_edge}
     \end{subfigure}
     \hfill
     \begin{subfigure}{0.48\textwidth}
         \centering
         \includegraphics[trim={180 345 180 350},clip,width=\textwidth]{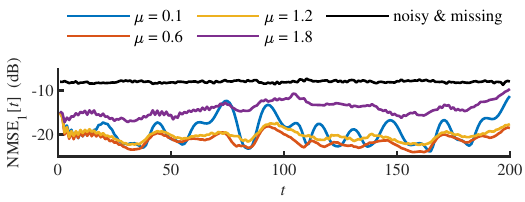}
         \caption{The total NMSE at each time point for all edge signals.}
         \label{fig_change_mu_NMSE}
     \end{subfigure}
     \caption{Performance of the ALMS-Hodge under fixed noise and different step sizes.}
     \label{fig_change_mu}
\end{figure}

\begin{figure}[htbp]
     \centering
     \begin{subfigure}{0.485\textwidth}
         \centering
         \includegraphics[trim={180 345 180 350},clip,width=\textwidth]{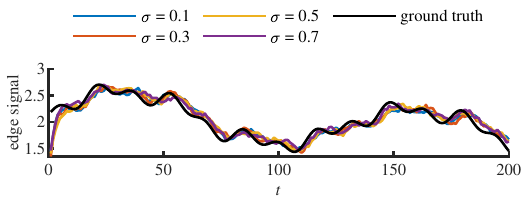}
         \caption{Reconstruction of an unobserved edge signal.}
         \label{fig_missing}
     \end{subfigure}
     \hfill
     \begin{subfigure}{0.485\textwidth}
         \centering
         \includegraphics[trim={180 345 180 350},clip,width=\textwidth]{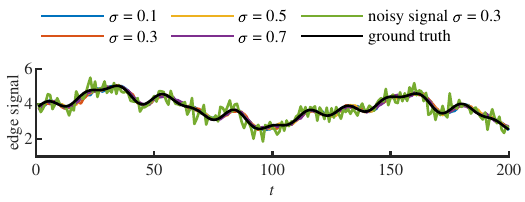}
         \caption{Denoising of an observed noisy signal.}
         \label{fig_observed}
     \end{subfigure}
     \caption{Performance of the ALMS-Hodge under fixed step size and different noise levels.}
     \label{fig_change_noise}
\end{figure}

\subsection{ALMS-Hodge analysis}
\label{sec_Hodge_exp}
Since the ALMS-Hodge is a component within AJVEE, we first conduct a few experiments to assess the performance of our ALMS-Hodge procedure in \eqref{eq_update_2} at estimating a time-varying edge signal in an online procedure on the Sioux Falls network. 
To specify \eqref{eq_update_2} onto the edges, we will be setting $k=1$, resulting in the edge signal estimation formula shown in \eqref{eq_AJVEE_k}.
The ALMS-Hodge is tested under the following two experiment settings. 
First, fix the noise level and run the ALMS-Hodge for 4 different step sizes.
Second, fix the step size and run the ALMS-Hodge for 4 different noise levels. 

In the experiment, we will use the low-pass filter shown in \eqref{eq_lp} with a cutoff frequency of $c = 0.5N_1$ to obtain the weights $\hat{\theta}_{1,p}$ in \eqref{eq_AJVEE_k} with $P = 7$. 
The experiments in this section are repeated 10 times; the averaged edge signal estimations of a few selected edges will be demonstrated along with the NMSE$_1[t]$ for all $N_1$ edge signals. 

For the fixed noise experiment, the noise is Gaussian distribution with mean $=0$ and variance $=0.1$; four different step sizes are $\mu_1 = 0.1, 0.6, 1.2$, and $1.8$. 
The signal reconstruction of the signal on one edge and the NMSE on the entire topology is shown in Figure.~\ref{fig_change_mu_edge} and Figure.~\ref{fig_change_mu_NMSE} respectively. 
The NMSE of comparing the ground truth against the noisy containing missing edge signal observations is included in Figure~\ref{fig_change_mu_edge} and denoted as noisy$\&$missing.
% The signal reconstruction of one missing edge and the denoising of one observed edge signal are shown in Fig~\ref{fig_change_mu} and Fig~\ref{fig_MSE_change_mu} respectively. 
From the edge signal estimation in Figure.~\ref{fig_change_mu_edge}, we can see that the step size parameter $\mu_1$ indeed controls the magnitude of the update. 
For $\mu_1 = 0.1$, the magnitude of the update is too small, which is reflected by visually inspecting Figure.~\ref{fig_change_mu_edge}: the estimation does not catch up with the time-varying change in the signal. 
As for the case when $\mu_1$ is too large ($\mu_1 = 1.8$), the large magnitude causes the estimation to fluctuate and unstable, which can be observed from Figure.~\ref{fig_change_mu_edge} as well. 
This confirms that $\mu_1$ controls the amount of update of the ALMS-Hodge.
The error of the estimation will be large when $\mu_1$ is either too large or too small. 
Inspecting the NMSE in Figure.~\ref{fig_change_mu_NMSE}, we can gain some insights on the design choice of $\mu_1$.

Now, we will be fixing the step size $\mu_1 = 0.6$ and conducting an online estimation of a time-varying edge signal under four different noise settings.
The Gaussian noise all have zero mean but four different variances $\sigma = 0.1, 0.3, 0.5$, and $0.7$.
The signal reconstruction of one missing edge is shown in Fig~\ref{fig_missing} and the denoising of one observed edge signal is shown in Fig~\ref{fig_observed}.
Upon examination of Figure.~\ref{fig_change_noise}, we verified that the ALMS-Hodge can effectively conduct online estimations of time-varying signals on the graph edges under Gaussian noise with various noise levels. 

\subsection{Joint Estimation on transportation network}
Returning to the question of joint estimation, we will be testing AJVEE based on Algorithm~\ref{algorithm_joint_node_edge} at jointly estimating time-varying vertex and edge signals by utilizing AJVEE within the ALMS-Hodge on the Sioux Falls network and the Anaheim network. 
% The previous two experiments have shown that the ALMS-Hodge could be applied to online simplicial estimation tasks. 
% The edge signal generation and unobserved data masking follow the same procedure in \ref{sec_Hodge_exp}. 

To show the advantages of using a TSP approach (the ALMS-Hodge) in AJVEE, we created a comparison algorithm that uses Line Graph LMS \cite{yan_2023_LGLMS} instead of using the ALMS-Hodge in the AJVEE.
We denote this approach as LGLMS in later experiments.
Different from the TSP-based ALMS-Hodge, the LGLMS uses line graphs to map edge signals on the vertices then processed by GSP. 
The aggregations terms used by AJVEE in this experiment are \eqref{eq_lower_adjacent_aggregation} for edge signals and \eqref{eq_upper_adjacent_aggregation} for vertex signals. 
Since the signal aggregation terms are missing in the LGLMS, we added the same signal aggregation terms that we use in AJVEE to LGLMS.
For edge signal estimation baselines, in addition to the LGLMS algorithm, we also included a non-adaptive TSP baseline using a basic low pass simplicial filter $\mathbf{\Sigma}_{1, \mathcal{F}_{lp}}$ similar to \eqref{filter}: $\hat{\boldsymbol{x}}_1[t+1] = \mathbf{U}_1\mathbf{\Sigma}_{1, \mathcal{F}_{lp}}\mathbf{U}_1^T\hat{\boldsymbol{x}}_1[t]$.
The last edge signal baseline is a Moving Average (MA) of the past 5 observations.

The vertex estimation results of AJVEE will be compared with the following three GSP baselines: the GLMS algorithm \cite{bib_LMS}, the adaptive Graph Least Mean $p^{th}$ algorithm (GLMP) \cite{bib_LMP}, and the adaptive Graph-Sign (G-Sign) algorithm\cite{yan_2022_sign}. 
Note that by the algorithm definitions in their original literature, the GSP methods are using a fixed graph Laplacian, so among all the baselines only the LGLMS algorithm is aware that the time-varying edge signals influence the vertices. 
In addition, we will feed the edge signal estimation of LGLMS into the vertex estimation scheme of the ALMS-Hodge as a comparison to purely using the ALMS-Hodge in AJVEE.
The MA is also included for the vertex signal estimation.

\begin{figure}[htbp]
     \centering
     \begin{subfigure}{0.48\textwidth}
         \centering
     \end{subfigure}
     \begin{subfigure}{0.48\textwidth}
         \centering
         \includegraphics[trim={180 345 180 340},clip,width=\textwidth]{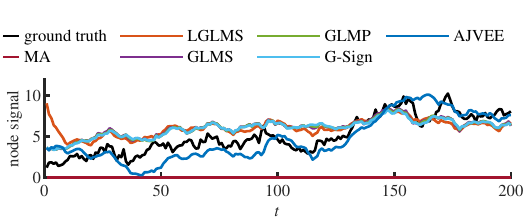}
          \includegraphics[trim={180 345 180 340},clip,width=\textwidth]{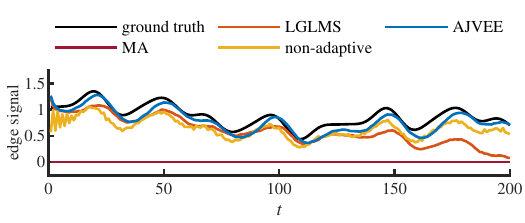} 
     \end{subfigure}
        \caption{Estimation on one of the unobserved edges (top) and one of the unobserved vertices (bottom) of the Sioux Falls network.}
        \label{fig_joint}
\end{figure}

\begin{figure}[htbp]
  \centering
       \centering
     \begin{subfigure}{0.48\textwidth}
         \centering
         \includegraphics[trim={180 330 180 340},clip,width=\textwidth]{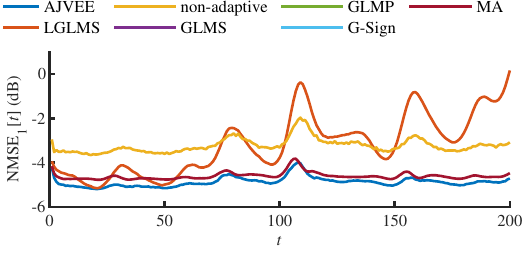}
     \end{subfigure}
     \begin{subfigure}{0.48\textwidth}
         \centering
         \includegraphics[trim={180 345 180 335},clip,width=\textwidth]{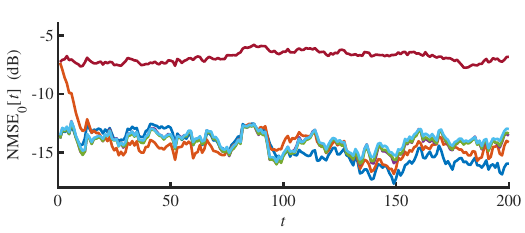}
     \end{subfigure}
\caption{NMSE of the AJVEE compared against separate estimation in the Sioux Falls network. Top: edges. Bottom: vertices.}
\label{fig_sioux_falls}
\end{figure}

\begin{figure}[htbp]
  \centering
       \centering
     \begin{subfigure}{0.48\textwidth}
         \centering
         \includegraphics[trim={180 330 180 335},clip,width=\textwidth]{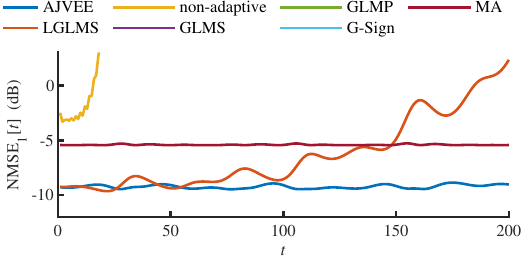}
     \end{subfigure}
     \begin{subfigure}{0.48\textwidth}
         \centering
         \includegraphics[trim={180 345 180 345},clip,width=\textwidth]{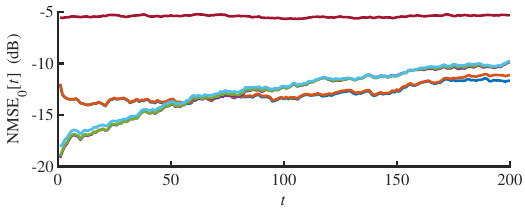}
     \end{subfigure}
\caption{NMSE of the AJVEE compared against separate estimation in the Anaheim network. Top: edges (the average of NMSE$_1[t]$ is the dashed line). Bottom: vertices.}
\label{fig_anaheim}
\end{figure}

The step sizes $\mu_k$ and the aggregation weights $r_{k}$ are tuned using grid search.
As for the filter used in the experiment, we again use an ideal low-pass filter defined by \eqref{eq_lp} on both the vertices and the edges.
The final parameter selections for AJVEE are shown in Table~\ref{table_parameter}. 
To promote fair comparison, all the tested algorithms share the same $\mu_0$ for vertex signal estimation and all the tested algorithms share the same $\mu_1$ for edge signal estimation.
Similar principle applies to the AJVEE and LGLMS for the aggregation weights.
The NMSE of the signal estimation results over the 100 runs are recorded in Fig.~\ref{fig_sioux_falls} for the Sioux Falls network and in Fig.~\ref{fig_anaheim} for the Anaheim network.
Additionally, an illustration of the estimation results on one of the unobserved edges and one of the unobserved vertices in the Sioux Falls network is shown in Fig.~\ref{fig_joint}. 

 
From Fig.~\ref{fig_sioux_falls} and Fig.~\ref{fig_anaheim}, AJVEE has lower NMSE than the baselines for the majority of the time points for the edge estimation task.
From the low NMSE, we can confirm that the AJVEE is accurate at tracking the changes of time-varying edge signals. 
AJVEE outperforms the LGLMS algorithm because the LGLMS does not operate directly on the edge signal; the LGLMS algorithm treats the original edge signal as vertex signals on a line graph projection. 
However, the projected signal in LGLMS is not guaranteed to be smooth on the $0$-simplices, so using the GSP nodal operations to predict edge signals based on a smooth graph assumption and a low-pass filter on the line graph projection does not perform well \cite{Schaub_2018}. 
As for the TSP non-adaptive method, the filtering operation $\mathbf{U}_1\mathbf{\Sigma}_{1, \mathcal{F}_{lp}}\mathbf{U}_1^T\boldsymbol{y}_1[t]$ is not optimized to reduce time-dependent errors compared to the adaptive update $\hat{\mathbf{H}}_1\boldsymbol{e}_1[t]$, which lead to the less compelling results compared to AJVEE. 
The MA has a relatively high NMSE because it can not make predictions on the missing edges, which can be observed from Figure~\ref{fig_joint} that the MA consistently outputs 0 for unobserved signals.
The other graph algorithms can infer the missing edge signals because each of the updates implies spatial diffusion, which means that these algorithms conduct spatial missing signal imputation and temporal 1-step predictions at the same time.
\begin{table}[htbp]
\centering
 \caption{Parameters settings}
\begin{tabular}{c c c c c c} 
% \hline
  Vertex Parameters & $\mu_0$ & $r_{0,\mathcal{S}}$& $r_{0,\bar{\mathcal{S}}}$ & $\mathcal{F}_0$ & $P$\\ 
   \hline
\hline
 Sioux Falls & 1.25 & 0.0025 & 0.05 & $[0, 0.4\lambda_{0,max}]$ & 7\\ 
 \hline
 Anaheim & 1.1 & 0.0001 & 0.00001 & $[0, 0.4\lambda_{0,max}]$ & 7\\ 
 \hline
\vspace{10 pt}
\end{tabular}
\centering
\begin{tabular}{c c c c c c} 
% \hline
   Edge Parameters & $\mu_1$ & $r_{1,\mathcal{S}}$ & $r_{1,\bar{\mathcal{S}}}$& $\mathcal{F}_1$ & $P$\\ 
\hline
 \hline
 Sioux Falls & 0.45 & 0.0025 & 0.15 & $[0, 0.58\lambda_{1,max}]$ & 7
\\ 
 \hline
 Anaheim & 0.75 & 0.00025 & 0.0005 & $[0, 0.58\lambda_{1,max}]$ & 7\\ 
 \hline
\end{tabular}
 \label{table_parameter}
\end{table}

The AJVEE has the lowest estimation error for the vertex signal estimation shown in Fig.~\ref{fig_sioux_falls} and Fig.~\ref{fig_anaheim}. 
The GSP approaches do not quite match the performance of AJVEE in predicting the time-varying vertex signals because the GSP methods used a fixed graph Laplacian to process the signals on the vertices.
Here, notice that the LGLMS performs only slightly worse than AJVEE.
The reason behind this is because, first, here in the vertex estimation the LGLMS is actually feeding the edge estimation results of LGLMS into the vertex update portion of the AJVEE, so here the LGLMS is essentially another variant of AJVEE. 
Second, the LGLMS estimation also benefited from the aggregation terms derived for ALMS-Hodge.
On the other hand, AJVEE utilizes the tive-varying edge signal to represent the diffusion dynamics of the vertex signals, which is more accurate at tracking the temporal evolving patterns of the time-varying vertex signals.
Our proposed AJVEE framework in Algorithm~\ref{algorithm_joint_node_edge} has demonstrated the ability to effectively capture the time-varying edge signals jointly with the time-varying vertex signals. 
LGLMS takes a similar approach as AJVEE to use edge signals to represent the time-varying regression, but the relatively inaccurate edge signal estimation causes the LGLMS to have inaccurate vertex signal estimation.
\begin{figure*}[htb]
     \centering
     \begin{subfigure}{0.3\textwidth}
         \centering
         \includegraphics[width=\textwidth]{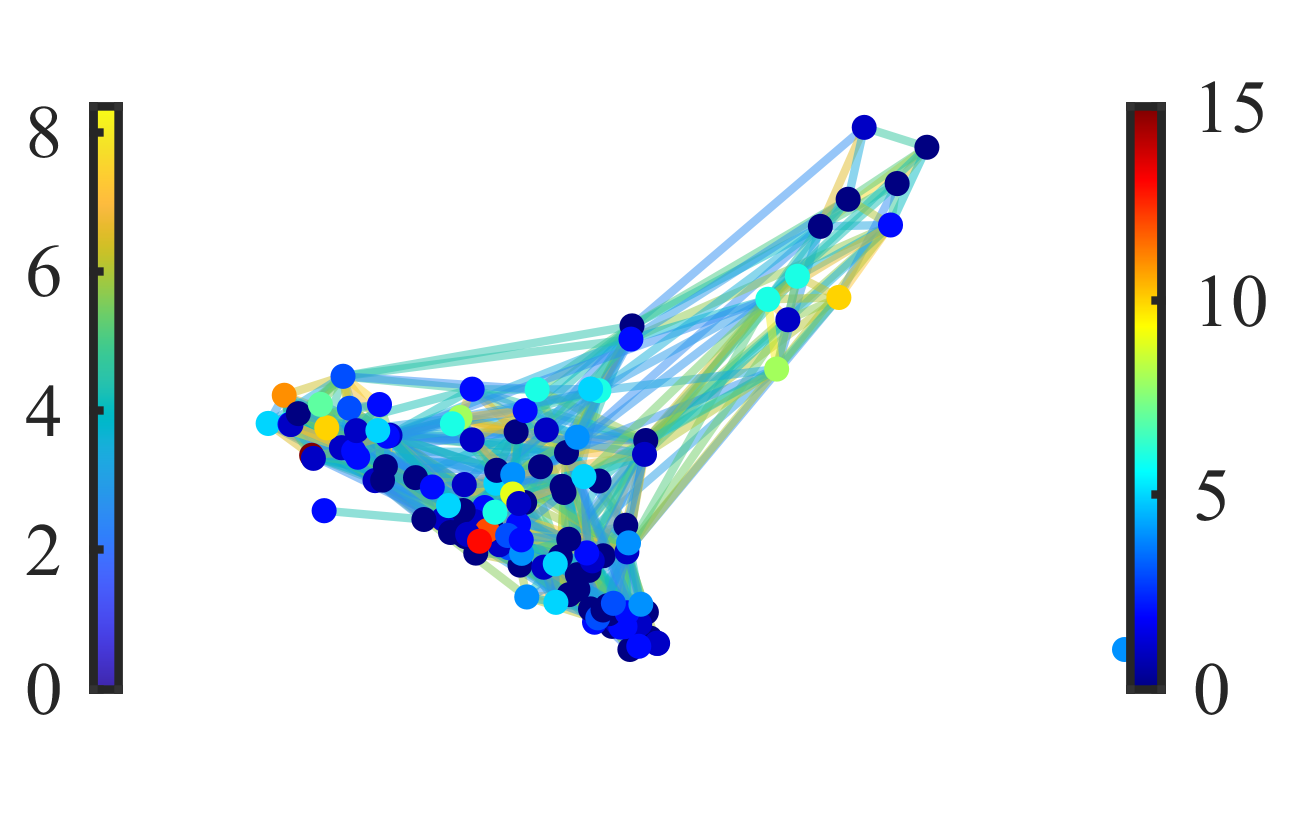}  
                  \vspace{-25pt}
     \end{subfigure}
      \begin{subfigure}{0.3\textwidth}
         \centering
         \includegraphics[width=\textwidth]{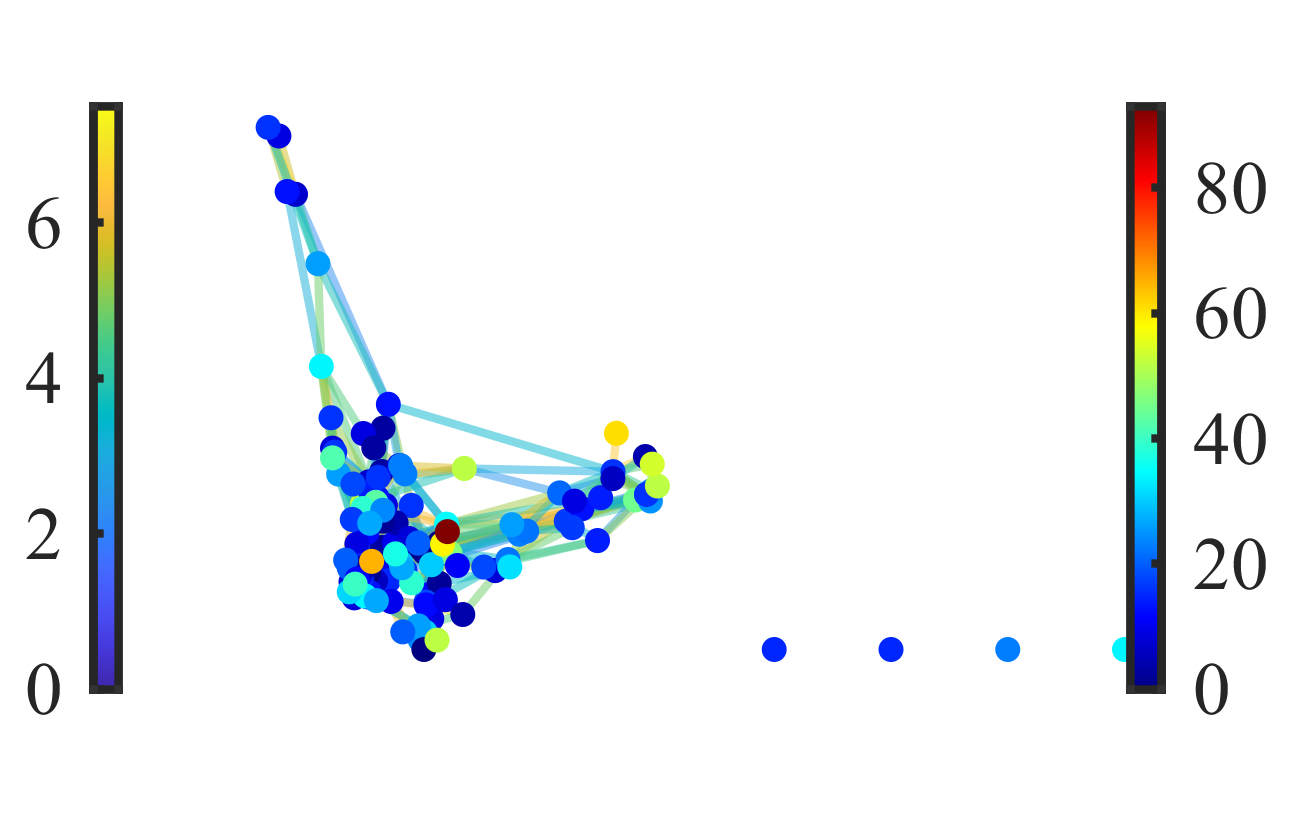}
                  \vspace{-25pt}
     \end{subfigure}
     \begin{subfigure}{0.3\textwidth}
         \centering
         \includegraphics[width=\textwidth]{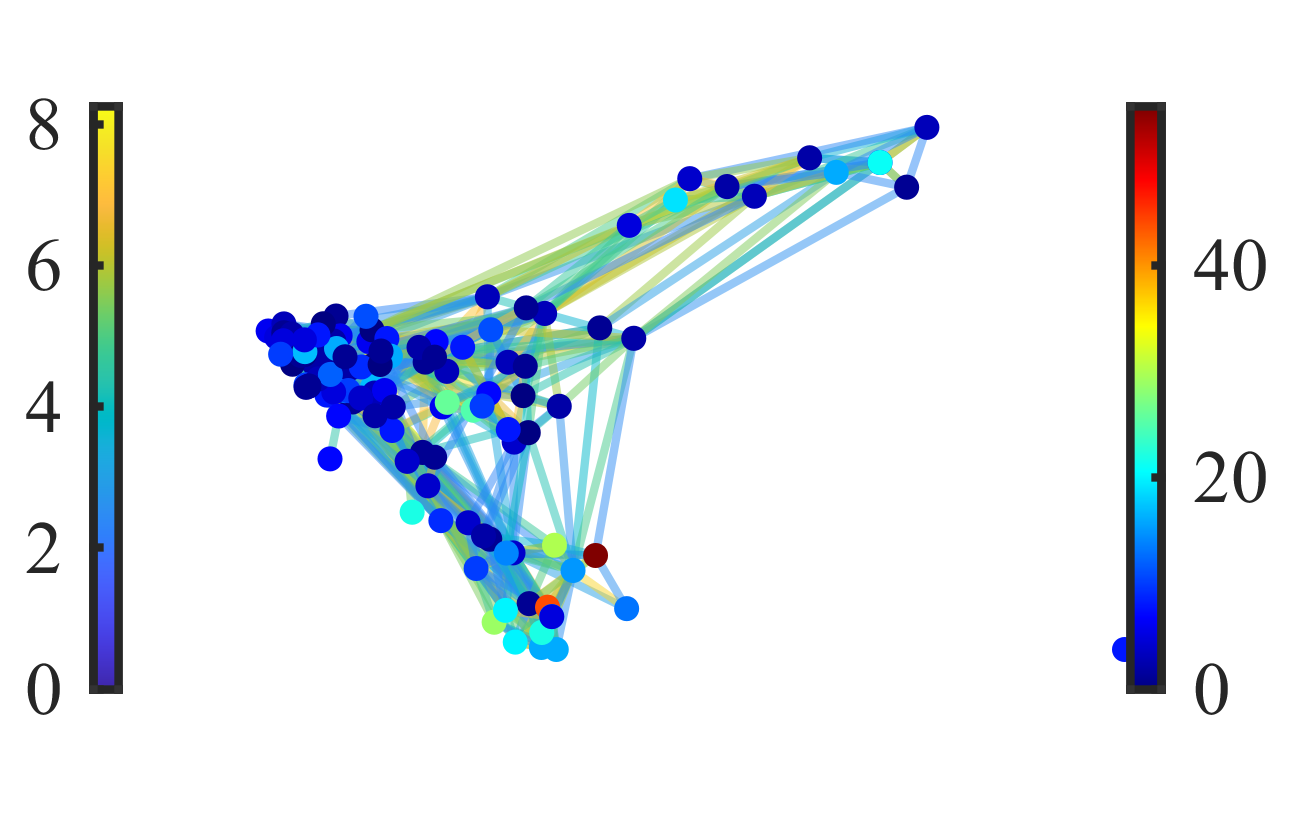}
                  \vspace{-25pt}
     \end{subfigure}
     \begin{subfigure}{0.3\textwidth}
         \centering
         \includegraphics[width=\textwidth]{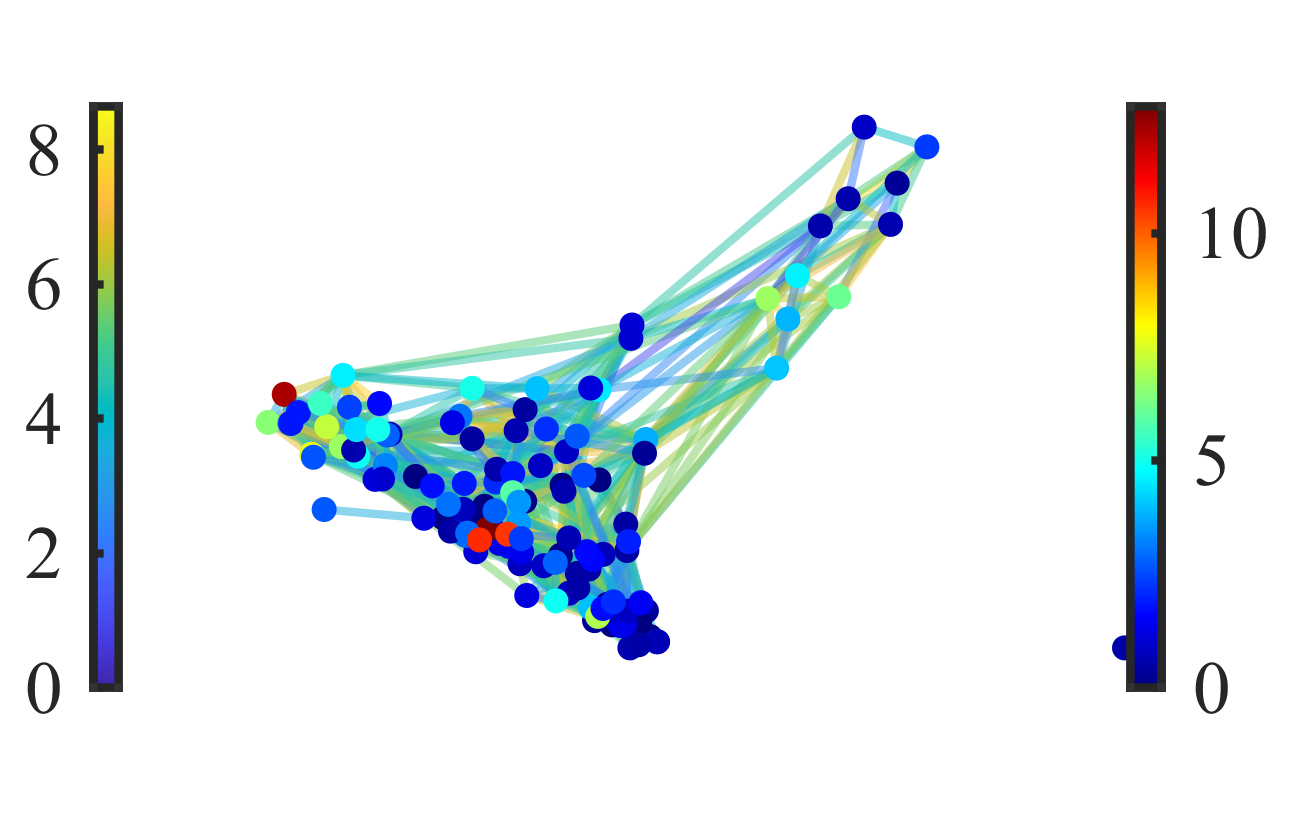}
         \vspace{-30pt}
         \caption*{$t=3$}        
     \end{subfigure}
      \begin{subfigure}{0.3\textwidth}
         \centering
         \includegraphics[width=\textwidth]{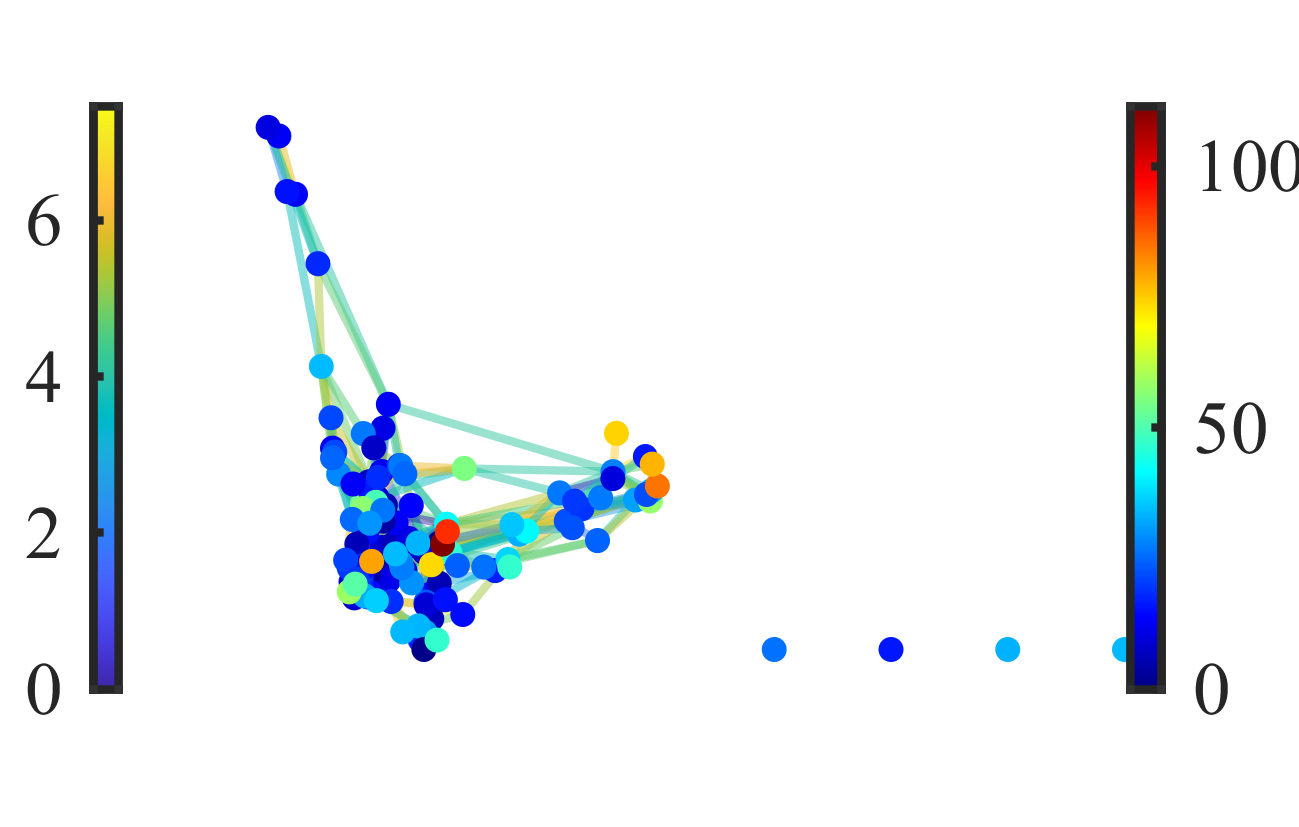}
                  \vspace{-30pt}
         \caption*{$t=30$}
     \end{subfigure}
     \begin{subfigure}{0.3\textwidth}
         \centering
         \includegraphics[width=\textwidth]{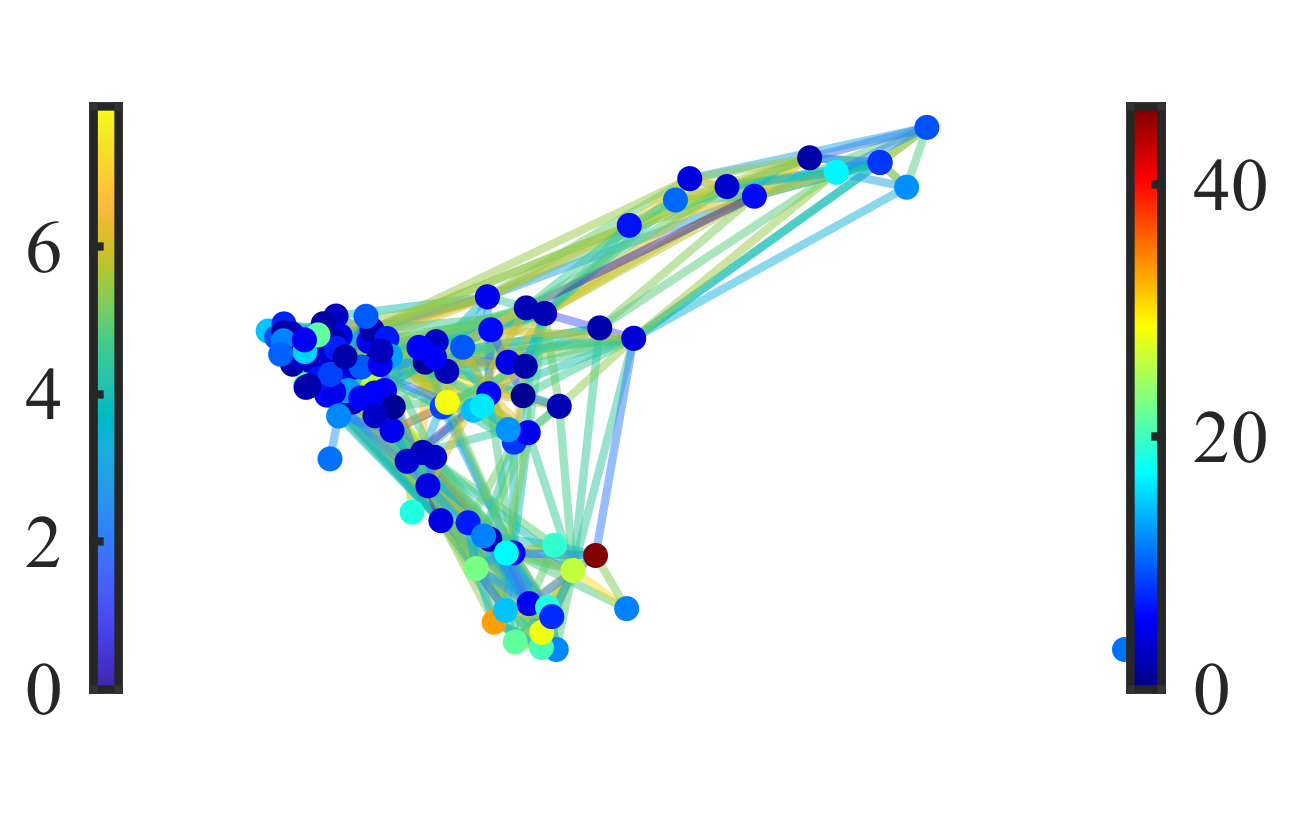}
                  \vspace{-30pt}
         \caption*{$t=61$}
     \end{subfigure}
        \caption{Dynamic graphs that record COVID-19 cases on the vertices and population mobility on the edges (self-loops are omitted). Upper row: ground truth. Lower row: predicted signals. Left color bar: edges (log scale). Right color bar: vertices.}
        \label{Covid_figs}
\end{figure*}

\subsection{COVID-19 case forecasting in population mobility network}
\label{sec_covid}
In this last experiment, we will use AJVEE to jointly estimate real-world time-varying vertex and edge signals in the England COVID-19 dataset using the procedure in Algorithm~\ref{algorithm_joint_node_edge}. 
The goal of this real-world dataset is to predict the confirmed number of COVID-19 cases in the future using the currently given information. 
It has been shown in \cite{Panagopoulos_2021} that in the England COVID-19 dataset, the number of confirmed cases (vertex signals) is positively correlated to the population mobility (edge signals), which indicates the validity of using AJVEE to conduct the online estimation. 
This means that our AJVEE framework is suitable to be applied to conduct online forecasting of the next-day cases as a diffusion modeled by a low-pass filter matches the diffusive nature of the data. 
Instead of the offline approaches proposed in \cite{Panagopoulos_2021}, we will take an online approach to this problem and make a one-day-ahead forecast of the number of cases.

For the England COVID-19 dataset, we will assume that the underlying topology is known when a new time instance arrives, but the simplicial signals on both the vertices and edges remain unknown and waiting to be predicted. 
During preprocessing, if multiple population mobility recordings exist between two vertices, we summed them up into a single edge and calculated the absolute sum of the population mobility. 
After the preprocessing, the population mobilities become signals on the edges and are not perceived using the concept of flows.
The edge signals are scaled to $1/1000$ of the original value in the preprocessing and are scaled back to the original scale after obtaining the output. 
We added Gaussian noise to both the ground truth vertex and edge signals. 
The dynamic nature of this dataset is modeled by a dynamic sampling matrix in the edges $\mathbf{D}_{1, \mathcal{S}}[t]$ where we compare the graph topology of the graph at the current time point $\mathcal{G}[t]$ with newly obtained graph topology $\mathcal{G}[t+1]$. 
If an edge is present in both $\mathcal{G}[t]$ and $\mathcal{G}[t+1]$, then the corresponding diagonal element in $\mathbf{D}_{1, \mathcal{S}}[t+1]$ will be $1$. 
If an edge is present in $\mathcal{G}[t+1]$ but not in $\mathcal{G}[t]$, then the corresponding diagonal element in $\mathbf{D}_{1, \mathcal{S}}[t+1]$ will be $0$. 
Both aggregations \eqref{eq_lower_adjacent_aggregation} and \eqref{eq_upper_adjacent_aggregation} are adopted in the experiment for predicting population mobility.
The predicted number of confirmed cases by AJVEE will be compared against three online GSP algorithms: GLMS \cite{bib_LMS}, GLMP \cite{bib_LMP}, and G-Sign \cite{yan_2022_sign}. 
All these GSP algorithms are modified to be aware of the topology change, but they are not provided with the time-varying population mobility on the graph edges.
The choice of this baseline is to demonstrate the usefulness of acquiring the population mobility at predicting the number of cases. 
We will also be using two simple baselines: 1. the MA of confirmed cases from the past 5 days; 2. repeat the noisy case count from the previous day (denoted as Last-day). 
This experiment reflects a real-world example of utilizing the time-varying signals on one simplicial dimension as an influence factor upon another simplicial dimension.

\begin{figure}[htbp]
    \centering
    \centerline{\includegraphics[trim={130 315 130 325},clip]{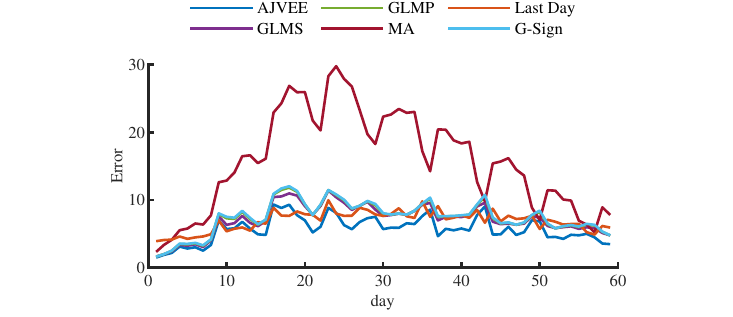}}
    \caption{Average prediction error per region of daily COVID-19 cases in England .}
    \label{fig_covid_node_error}
\end{figure}

The predicted population mobility and the predicted number of cases at three different time points using Algorithm~\ref{algorithm_joint_node_edge} are illustrated in Fig.~\ref{Covid_figs}. 
From Fig.~\ref{Covid_figs}, it is clear that the recovered population mobility and the number of cases are close to the ground truth.
In Fig.~\ref{fig_covid_node_error}, we calculated the average absolute error of the predicted number of cases at each time point in all the vertices, which is the suggested metric by the dataset \cite{Panagopoulos_2021}.
By inspecting Fig.~\ref{fig_covid_node_error}, we observe that our joint estimation framework has the lowest error most of the time. 
AJVEE outperforms GSP algorithms that purely use vertex information because the GSP algorithms are unable to modify the magnitude of the update based on the information from the population mobility.
However, because AJVEE jointly estimates the edge and vertex signals, AJVEE is more aware of the change in the vertices that are caused by the change in the edges.
The nonstationary behavior of the confirmed cases causes relatively worse performance of MA and Last-day.
The results in this experiment indicate that our proposed AJVEE framework could utilize and jointly predict the dynamic information on the edges and the vertices to solve a real-world problem of online prediction of time-varying signals. 
The current approach to this real-world problem is a bit primitive. 
In the next section, we will discuss potential improvements to our AJVEE framework to solve similar problems.

\section{Conclusion, Limitation, and Future Work}
% \section{Conclusion}
\label{sec_conclusion}
This paper presented the AJVEE framework for estimating time-varying vertex signals and edge signals, in which the dynamics of the vertex signals are in the form of time-varying edge signals. 
The underlying online estimation method of AJVEE is the ALMS-Hodge, an adaptive procedure backed by spatial and spectral TSP techniques. 
By treating graphs as a special case of a simplicial complex, two ALMS-Hodge with different simplicial orders are specified on the vertices and edges and then merged into the joint estimation. 
Even though AJVEE is proposed on the vertices and edges, we also discussed a more generalized setting where AJVEE is extended to higher-order structures.
In summary, AJVEE effectively addresses the challenges of representing topological irregularity, estimating unknown data from known data, removing noise present in the data, and extracting cross-space-time variations.
Experiments on both synthetic and real data indicate that AJVEE can efficiently solve the task of joint online estimation of time-varying vertex and edge signals. 

Currently, one limitation of AJVEE is that the influence of edge signals is applied globally to all vertex signals using operations defined by the Hodge-Laplacian, without considering the direction. 
Even though the sign and directions at the time of edge estimation, when casting the edge signals as the influence of the vertex signals, the direction and sign information are not properly utilized, leading to an invariant model.
However, these directional information and sign differences in edge signals, which can change over time, can be important to the data. 
The current invariant treatment of edge signals can lead to sub-optimal performance in downstream tasks that depend on these directional and sign attributes. 
We recognize this as a challenge and suggest that future work should explore equivariant models to better account for these variations, rather than relying on the current invariant model.
Another limitation in our current setup is that we are not able to process directed multi-edges between nodes. 
Additionally, the underlying topology is assumed to be known. 
One possible direction worth exploring is combining our AJVEE framework with topology learning algorithms. 
This combination will further extend the predictive power of AJVEE onto graphs that dynamically evolve the topology over time.
The extension of TSP approaches to dynamic graphs, multi-graphs, and equivariant models will be interesting future directions. 
Due to the lack of suitable real-world time-varying traffic flows for our experimental setting, we utilized a combination of synthetic data generated from real data on top of a graph representing real-world road networks. 
This is a limitation in the current experiment design and we look forward to applying our AJVEE framework to additional real-world datasets that contain real time-varying features.

%\appendices
%\section{appendices}

% % use section* for acknowledgment
% \section*{Acknowledgment}
% We would like to express our deep gratitude to Professor Xin Guo from the  Department of Industrial Engineering and Operations Research, University of California, Berkeley. Thank you for your hospitality during Yi Yan's visit. The kind support from your group greatly facilitated this research. 

\bibliographystyle{IEEEbib}
\bibliography{paper}

\appendices
\section{ALMS-Hodge Convergence analysis}
\label{app_convergence}
To make sure the ALMS-Hodge procedure in the AJVEE outputs a stable estimation, we analyze the update of the ALMS-Hodge under steady-state simplicial signal estimation. 
For ease of analysis, we will not be using the polynomial approximation of $\mathbf{H}_k$ shown in \eqref{eq_cheb_H_k} in this analysis but rather use the spectral definition $\mathbf{H}_k =  \mathbf{U}_{k,\mathcal{F}}(\mathbf{U}_{k,\mathcal{F}}^T\mathbf{D}_{k, \mathcal{S}}\mathbf{U}_{k,\mathcal{F}})^{-1}\mathbf{U}_{k,\mathcal{F}}^T$. 
Suppose that we have a clean simplicial signal $\boldsymbol{x}_k$, the update error $\tilde{\boldsymbol{x}}_k[t]$ at time $t$ is the difference between the current estimation $\hat{\boldsymbol{x}}_k[t]$ and $\boldsymbol{x}_k$: $\tilde{\boldsymbol{x}}_k[t] = \boldsymbol{x}_k-\hat{\boldsymbol{x}}_k[t]$. 
The ALMS-Hodge is stable if the update error converges under steady state estimation. 
To begin the proof, we use the SFT to transfer the update function \eqref{eq_update_2} in the spectral domain:
\begin{equation}
    \tilde{\boldsymbol{s}}_k[t+1] = \tilde{\boldsymbol{s}}_k[t]+\mu_k\mathbf{U}_{k,\mathcal{F}}^T\mathbf{H}_k\boldsymbol{e}_k[t]+r_k\mathbf{U}_{k,\mathcal{F}}^T\mathbf{R}_k\mathbf{U}_{k,\mathcal{F}}\hat{\boldsymbol{s}}_k[t],
    \label{update_error_1}
\end{equation}

Since the aggregation term is recursive based on $\hat{\mathbf{s}}_k[t]$, we will first prove the update error $\tilde{\boldsymbol{s}}_k[t]$ converges independent of the aggregation term. 
The estimation error $\boldsymbol{e}_k[t]$ can be written in terms of $\tilde{\mathbf{s}}_k[t]$ and $\mathbf{s}_k$ in the spectral domain as $\boldsymbol{e}_k[t] = \mathbf{D}_\mathcal{S}\mathbf{U}_{k,\mathcal{F}}(\boldsymbol{s}_k+\boldsymbol{\eta}_k[t]-\hat{\boldsymbol{s}}_k[t])$. 
Then, dropping the aggregation, \eqref{update_error_1} is rearranged into
\begin{equation}
      \tilde{\boldsymbol{s}}_k[t+1] = \mathbf{\Omega}\tilde{\boldsymbol{s}}_k[t]  +\mu_k\mathbf{U}_{k,\mathcal{F}}^T\mathbf{H}_k\mathbf{D}_{k, \mathcal{S}}\boldsymbol{\eta}_k[t],  
      \label{update_error_2}
\end{equation}
where $\mathbf{\Omega} = (\mathbf{I}-\mu_k\mathbf{U}_{k,\mathcal{F}}^T\mathbf{H}_k\mathbf{D}_{k, \mathcal{S}}\mathbf{U}_{k,\mathcal{F}})$. 
Equation \eqref{update_error_2} can be expressed as a recursive update starting from $t=0$:
\begin{equation}
        \tilde{\boldsymbol{s}}_k[t+1] = \mathbf{\Omega}^i\tilde{\boldsymbol{s}}_k[0]
        +\mu_k\sum_{i = 1}^{t}\mathbf{\Omega}^{i-1}\mathbf{U}_{k,\mathcal{F}}^T\mathbf{H}_k\mathbf{D}_{k, \mathcal{S}}\boldsymbol{\eta}_k[t].
              \label{update_error_3}
\end{equation}
Now, we can find the update error in the mean squared sense by calculating the variance of \eqref{update_error_3}: 
\begin{equation}
\begin{split}
    &\mathbb{E}\|\tilde{\boldsymbol{s}}_k[t+1]\|^2 = \mathbb{E}\|\tilde{\boldsymbol{s}}_k[0]\|^2_{\mathbf{\Phi}^t}+\\
    &\mu_k^2\mathbb{E}\sum_{i = 0}^{t-1}(\mathbf{U}_{k,\mathcal{F}}^T\mathbf{H}_k\mathbf{D}_{k, \mathcal{S}}\boldsymbol{\eta}_k[t])^T\mathbf{\Phi}^{i}(\mathbf{U}_{k,\mathcal{F}}^T\mathbf{H}_k\mathbf{D}_{k, \mathcal{S}}\boldsymbol{\eta}_k[t])^2.
    \end{split}
    \label{update_error_4}
\end{equation}
For notation simplicity, we set $\mathbf{\Phi} = \mathbf{\Omega}^T\mathbf{\Omega}$ and use the weighted Euclidean norm $\|\tilde{\boldsymbol{s}}_k[0]\|_{\mathbf{\Phi}^t} = \|\tilde{\boldsymbol{s}}_k[0]^T\mathbf{\Phi}^t\tilde{\boldsymbol{s}}_k[0]\|$ in \eqref{update_error_4}.
Then, we rearrange some terms by using the Trace trick $\mathbb{E}\left(\mathbf{X}^T\mathbf{YX}\right) = $ Tr$\left(\mathbb{E}\left(\mathbf{XX}^T\mathbf{Y}\right)\right)$, leading to
\begin{equation}
         \mathbb{E}\|\tilde{\boldsymbol{s}}_k[t+1]\|^2 = \mathbb{E}\|\tilde{\boldsymbol{s}}_k[0]\|^2_{\mathbf{\Phi}^t}+\mu_k^2\text{Tr}\left(\sum_{i = 0}^{t-1}\mathbf{\Psi}\mathbf{\Phi}^i\right),
         \label{update_error_5}
\end{equation}
where $\mathbf{\Psi} = \mathbf{U}_{k,\mathcal{F}}^T\mathbf{H}_k\mathbf{D}_{k, \mathcal{S}}\mathbf{C}_k\mathbf{H}_k^T\mathbf{U}_{k,\mathcal{F}}$ and $\mathbf{C} = \text{cov}(\boldsymbol{\eta},\boldsymbol{\eta})$. 
With some further derivations using two properties Tr$\left(\mathbf{YX}\right) = $ vec$(\mathbf{X}^T)$vec$(\mathbf{Y})$ and vec$(\mathbf{XYZ}) = (\mathbf{Z}^T\otimes\mathbf{X})$vec$(\mathbf{Y})$, we can factor out the terms independent of the summation index $i$ from \eqref{update_error_5}.
The summation in \eqref{update_error_5} becomes a geometric series when the condition $\|\mathbf{\Omega}\|<1$ is satisfied, then
\begin{equation}
         \mathbb{E}\|\tilde{\boldsymbol{s}}_k[t+1]\|^2 = \mathbb{E}\|\tilde{\boldsymbol{s}}_k[0]\|^2_{\mathbf{\Phi}^t}+
    \mu_k^2\text{vec}(\mathbf{\Psi})^T\sum_{i = 0}^{t-1}\mathbf{Q}^i\text{vec}(\mathbf{I}),
    \label{update_error_6}
\end{equation}
where $\mathbf{Q} = \mathbf{\Omega}^T\otimes\mathbf{\Omega}$. 
Taking the limit with $\lim_{t\to\infty} \mathbb{E}\|\tilde{\boldsymbol{s}}_k[t+1]\|^2$ and combining the condition of  $\|\mathbf{\Omega}\|<1$, we recognize the right side of \eqref{update_error_6} converges to a constant. 
The only user-defined parameter in the ALMS-Hodge without aggregation is a positive step size $\mu_k$.
Using the properties of $l_2$-norm, we can have an inequality $\|\mathbf{\Omega}\|<\|\mathbf{I}\|+\mu_k\|(\mathbf{U}_{k,\mathcal{F}}^T\mathbf{H}_k\mathbf{D}_{k, \mathcal{S}}\mathbf{U}_{k,\mathcal{F}}\|<1$. 
Now, because the sampling matrix $\mathbf{\mathbf{D}_{k, \mathcal{S}}}$ is idempotent and self-adjoint, the matrix $\mathbf{U}_{k,\mathcal{F}}^T\mathbf{H}_k\mathbf{D}_{k, \mathcal{S}}\mathbf{U}_{k,\mathcal{F}}$ is symmetric, and its $l_2$-norm will be less or equal to its largest eigenvalue. 
This means that the ALMS-Hodge has a bounded update error and the algorithm converges if $\mu_k$ is chosen under the condition
\begin{equation}
    0<\mu_k<\frac{2}{\lambda_{\text{max}}(\mathbf{U}_{k,\mathcal{F}}^T\mathbf{H}_k\mathbf{D}_{k, \mathcal{S}}\mathbf{U}_{k,\mathcal{F}})},
    \label{converge_condition_1}
\end{equation}
where $\lambda_{\text{max}}()$ calculates the maximum eigenvalue of a matrix. 
Now, with the selection condition of $\mu_k$ provided, let us look back into \eqref{update_error_6}. 
If we choose a smaller value for $\mu_k$, the term $\mathbf{\Omega}$ will be large, so the squared error will converge to a smaller value but will take more iterations to converge. 
On the other hand, if we choose a larger value for $\mu_k$, $\mathbf{\Omega}$ is small, and the algorithm converges faster but with higher error.

The ALMS-Hodge converges with the aggregation term. 
To see this, an expression similar to \eqref{update_error_2} but with the aggregation can be formulated as 
\begin{equation}
    \begin{split}
              \tilde{\boldsymbol{s}}_k[t+1] = \mathbf{\Omega}_r\tilde{\boldsymbol{s}}_k[t] +\mu_k\mathbf{U}_{k,\mathcal{F}}^T\mathbf{H}_k\mathbf{D}_{k, \mathcal{S}}\boldsymbol{\eta}_k[t]\\
              +r\mathbf{U}_{k,\mathcal{F}}^T\mathbf{R}_k\mathbf{U}_{k,\mathcal{F}}\boldsymbol{s}_k,  
    \end{split}
\end{equation}
where $\boldsymbol{s}_k = \mathbf{U}_{k,\mathcal{F}}^T\boldsymbol{x}_k$ and $\mathbf{\Omega}_r = (\mathbf{\Omega}+r\mathbf{U}_{k,\mathcal{F}}^T\mathbf{R}_k\mathbf{U}_{k,\mathcal{F}})$. 
Let $\mathbf{\Phi}_r = \mathbf{\Omega}_r^T\mathbf{\Omega}_r$ and $\mathbf{Q}_r = \mathbf{\Omega}_r^T\otimes\mathbf{\Omega}_r$, then take the same derivation steps as before, the update error in the mean squared sense is 
\begin{equation}
    \begin{split}
        \mathbb{E}\|\tilde{\boldsymbol{s}}_k[t+1]\|^2 = &\mathbb{E}\|\tilde{\boldsymbol{s}}_k[0]\|^2_{\mathbf{\Phi}^t_r}+\mathbb{E}\|r\mathbf{U}_{k,\mathcal{F}}^T\mathbf{R}_k\mathbf{U}_{k,\mathcal{F}}\boldsymbol{s}_k\|^2\\
&+\mu_k^2\text{vec}(\mathbf{\Psi})^T\sum_{i = 0}^{t-1}\mathbf{Q}^i_r\text{vec}(\mathbf{I}).
    \end{split}   
    \label{update_error_7}
\end{equation}
The condition for  \eqref{update_error_7} to converge is again $\|\mathbf{\Omega}_r\|<1$. 
Now, using the property of the $l_2$-norm, we have the inequality $\|\mathbf{\Omega}_r\|<\|\mathbf{\Omega}\|+\|r\mathbf{U}_{k,\mathcal{F}}^T\mathbf{R}_k\mathbf{U}_{k,\mathcal{F}}\|$. 
To be on the safe side, and assuming that $r>0$, the ALMS-Hodge converges if \eqref{converge_condition_1} is met and if we have $\|\mathbf{\Omega}\|+r\|\mathbf{U}_{k,\mathcal{F}}^T\mathbf{R}_k\mathbf{U}_{k,\mathcal{F}}\|<1$. 
As a result, the range of choice of selecting parameter $r$ is 
\begin{equation}
    0<r<\frac{1-\mathbf{\Omega}}{\lambda_{\text{max}}(\mathbf{U}_{k,\mathcal{F}}^T\mathbf{R}_k\mathbf{U}_{k,\mathcal{F}})}.
    \label{converge_condition_2}
\end{equation}

\begin{figure}[htb]
  \centering
  \centerline{\includegraphics[trim={180 345 180 345},clip,width=8.8cm]{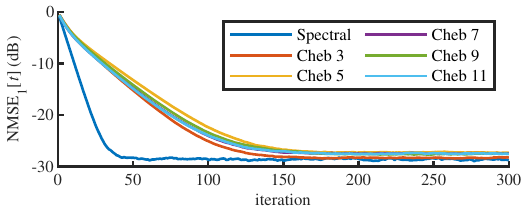}}
\caption{Steady-state estimation on the edges of the Sioux Falls network using ALMS-Hodge with different step sizes.}
\label{fig_muv}
\end{figure}

In order to observe the convergence behavior of the ALMS-Hodge, we will make online estimations for 1200 iterations where at each iteration we have a new noisy and missing observation of the steady-state edge signal. 
Gaussian distribution with mean $=0$ and variance $=0.18$ will be used for the noise distribution  $\boldsymbol{\eta}[t]$.
The step size is set to $\mu_1 = 0.1$.
The low pass filter used in the experiment has the pass band of 26 lowest frequencies.
Both spatial ALMS-Hodge with order P (denoted as Cheb P) and spectral ALMS-Hodge will be tested.
We will be testing the ALMS-Hodge for 6 different settings: spectral ALMS-Hodge, Cheb 3, Cheb 5, Cheb 7, Cheb 9, and Cheb 11. 
The NMSE of this steady-state experiment is shown in Fig.~\ref{fig_muv}.

By inspecting Fig~\ref{fig_muv}, we can observe that all variants of the ALMS-Hodge result in similar NMSE results after 200 iterations. 
Additionally, all the spatial update functions have similar NMSE performance to the spectral update function when converged. 

\section{Spectral analysis of AJVEE}
\label{app_spectral_analysis_AJVEE}
There are two incidence matrices we are particularly interested in for the estimation of node and edge signals, namely the node-to-edge incidence matrix is $\mathbf{B}_1$ and the edge-to-triangle incidence matrix is $\mathbf{B}_2$. 
The matrix $\mathbf{B}_1$ is a divergence operator because $\mathbf{B}_1\boldsymbol{x}_1[t] \in \mathbb{R}^{N_0}$ calculates the difference between the outgoing edge signals and the incoming edge signals at a vertex. 
An edge signal $\boldsymbol{x}_1[t]$ is considered divergence-free if $\mathbf{B}_1\boldsymbol{x}_1[t] = 0$ \cite{Jia_2019,Yang_2021_FIR}.  
The gradient operator $\mathbf{B}_1^T$ on the vertex signal $\boldsymbol{x}_0$ calculates the difference between the endpoints in each edge along the edge orientation, inducing a gradient flow $\boldsymbol{x}_G[t] = \mathbf{B}_1^T\boldsymbol{x}_0[t]  \in \mathbb{R}^{N_1}$ on the edges. 
The matrix $\mathbf{B}_2^T$ is a curl operator since $\mathbf{B}_2\boldsymbol{x}_1[t] \in \mathbb{R}^{N_1}$ computes the net flow of the edges along each triangle. 
If $\mathbf{B}_2^T\boldsymbol{x}_1 = 0$ holds true, then the edge signal $\boldsymbol{x}_1[t]$ is curl-free \cite{Yang_2022_Simplicial}. 
The curl adjoint operator $\mathbf{B}_2$ induces the curl flow onto the edges from the triangles by $\boldsymbol{x}_C[t] = \mathbf{B}_2\boldsymbol{x}_2[t]$.
From the Hodge decomposition, aggregations that utilize the four abovementioned operators can be defined using \eqref{eq_lower_adjacent_aggregation}, \eqref{eq_upper_adjacent_aggregation}, \eqref{eq_boundary_aggregation}. and \eqref{eq_coboundary_aggregation}.

Taking an edge signal $\boldsymbol{x}_1[t]$ as an example, in the spectral domain, the Hodge decomposition in \eqref{eq_hodge_decomposition} implies that we can modify the signal $\boldsymbol{x}_1[t]$ and the update error $\boldsymbol{e}_1[t]$ by modifying the harmonic component, the gradient component, and the curl component separately. 
Based on the Hodge decomposition, the eigenvector matrix $\mathbf{U}_1$ and the corresponding eigenvalue matrix $\mathbf{\Lambda_1}$ of $\mathbf{L}_1$ can be rearranged as shown below \cite{Yang_2022_Simplicial}:
\begin{equation}
                 \mathbf{U}_1 = [\mathbf{U}_H , \mathbf{U}_G, \mathbf{U}_C] \text{ and }
       \mathbf{\Lambda}_1 = \text{blkdiag}(\mathbf{\Lambda}_H, \mathbf{\Lambda}_G, \mathbf{\Lambda}_C),
            \label{eq_laplacian_decomposed}
\end{equation}
where blkdiag() is the block diagonal operation. 
The matrices $\mathbf{U}_H$, $\mathbf{U}_G$, and $\mathbf{U}_C$ in \eqref{eq_laplacian_decomposed} are the eigenvector matrix of the harmonic component, the gradient component, and the curl component respectively. 
The diagonal matrices $\mathbf{\Lambda}_H$, $\mathbf{\Lambda}_G$, and $\mathbf{\Lambda}_C$ contains the eigenvalue corresponds to the eigenvectors in $\mathbf{U}_H$, $\mathbf{U}_G$, and $\mathbf{U}_C$.
Because $\mathbf{L}_1$ can be represented as the sum of $\mathbf{L}_{1,l}$ and $\mathbf{L}_{1,u}$, the following expressions can be derived using the logic in \eqref{eq_laplacian_decomposed} \cite{Yang_2022_Simplicial}:
\begin{equation}
    \begin{split}
                 &\mathbf{L}_{1,l} = \mathbf{U}_1 \text{blkdiag}(\mathbf{0}, \mathbf{\Lambda}_G, \mathbf{0}) \mathbf{U}_1^T, \\
       &\mathbf{L}_{1,u}  = \mathbf{U}_1 \text{blkdiag}(\mathbf{0}, \mathbf{0}, \mathbf{\Lambda}_C) \mathbf{U}_1^T.
    \end{split}
            \label{eq_eq_laplacian_decomposed_2}
\end{equation}
Now, by looking at \eqref{eq_laplacian_decomposed} and \eqref{eq_eq_laplacian_decomposed_2}, it is apparent that by carefully designing the filter $\mathbf{\Sigma}_1$ in \eqref{H_k} in the spectral domain, we could achieve the direct modification of the harmonic component, the gradient component, and the curl component of the update $\boldsymbol{e}_1[t]$ as well as the estimated signal $\hat{\boldsymbol{x}}_1[t]$.  

A similar analysis can be conducted on the graph vertices by decomposing $\mathbf{L}_0$ or $\hat{\mathbf{L}}_0[t]$.
It should be pointed out that the vertices only have upper adjacency, which simplifies the spectral analysis. 
Generally speaking, in vertex signal estimation tasks the aggregation requires a smooth (low frequency) signal assumption \cite{Ortega_2018}. 
In the GSP perspective, if $\hat{\mathbf{H}}_0[t]$ is defined based on an approximation of a low-pass filter $\mathbf{\Sigma}_{\mathcal{F},0}$, then it enforces the estimated signal to be smooth \cite{stankovic_2019_vertex}. 
\end{document}